\documentclass[11pt,USenglish]{article}
\pdfoutput=1

\usepackage{setspace}
\usepackage{cite}

\renewcommand{\Phi}{\phi}

\usepackage[utf8]{inputenc}
\usepackage{geometry}
\usepackage[USenglish]{babel}
\usepackage{verbatim}
\usepackage{prettyref}
\usepackage{amsmath}
\usepackage[normalem]{ulem}
\usepackage{amssymb}
\usepackage{graphicx}
\usepackage{setspace}
\usepackage{esint}

\usepackage{color}
\definecolor{darkgreen}{rgb}{0,0.5,0}
\definecolor{darkblue}{rgb}{0,0,0.6}
\definecolor{purple}{rgb}{0.4,.2,0.7}
\definecolor{orange}{rgb}{0.95, 0.5, 0.3}

\usepackage[colorlinks=true,citecolor=darkgreen,linkcolor=darkblue,urlcolor=blue]{hyperref}

\numberwithin{equation}{section}
\numberwithin{figure}{section}
\numberwithin{table}{section}

\def\be{\begin{equation}}
\def\ee{\end{equation}}
\def\bea{\begin{eqnarray}}
\def\eea{\end{eqnarray}}
\def\ba{\begin{align}}
\def\ea{\end{align}}
\def\d{\text{d}}

\def\cO{{\cal O}}
\def\d{\text{d}}

\newcommand{\bz}{\bar{z}}

\begin{document}
\begin{spacing}{1.3}

~
\vskip5mm

\begin{center} {\Large \bf From Conformal Blocks to Path Integrals in the Vaidya Geometry}

\vskip10mm

Tarek Anous,$^{1}$ Thomas Hartman,$^{2}$ Antonin Rovai$^{3}$ \& Julian Sonner$^{3}$\\
\vskip1em
{\it 1) Department of Physics and Astronomy, University of British Columbia, 6224 Agricultural Road, Vancouver, B.C. V6T 1Z1, Canada} \\
\vskip5mm
{\it 2) Department of Physics, Cornell University, Ithaca, New York, USA} \\
\vskip5mm
{\it 3) Department of Theoretical Physics, University of Geneva, 24 quai Ernest-Ansermet, 1214 Gen\`eve 4, Switzerland}
\vskip5mm

\tt{ tarek@phas.ubc.ca, hartman@cornell.edu, \{antonin.rovai,julian.sonner\}@unige.ch}

\end{center}

\vskip10mm

\begin{abstract}

Correlators in conformal field theory are naturally organized as a sum over conformal blocks.  In holographic theories, this sum must reorganize into a path integral over bulk fields and geometries.  We explore how these two sums are related in the case of a point particle moving in the background of a 3d collapsing black hole. The conformal block expansion is recast as a sum over paths of the first-quantized particle moving in the bulk geometry.  Off-shell worldlines of the particle correspond to subdominant contributions in the Euclidean conformal block expansion, but these same operators must be included in order to correctly reproduce complex saddles in the Lorentzian theory. During thermalization, a complex saddle dominates under certain circumstances; in this case, the CFT correlator is not given by the Virasoro identity block in any channel, but can be recovered by summing heavy operators. This effectively converts the conformal block expansion in CFT from a sum over intermediate states to a sum over channels that mimics the bulk path integral.

\end{abstract}

\pagebreak
\pagestyle{plain}

\setcounter{tocdepth}{2}
{}
\vfill
\tableofcontents
\section{Introduction}

The AdS/CFT correspondence equates the bulk path integral to the CFT generating functional,
\be\label{dict}
\int_J D g \, D\phi \, \exp\left(i S_{\text{bulk}}[g,\phi]\right) = Z_{\textsc{cft}}[J] \ ,
\ee
where $g$ is the bulk metric and $\phi$ denotes all the other bulk fields. Boundary conditions in the bulk are set by sources $J$ in the CFT.  This expression is somewhat schematic non-perturbatively, since the path integral on the left is difficult to define any other way. It necessarily includes a sum over off-shell geometries, and off-shell configurations of all the bulk fields $\phi$.

Nonetheless in the semiclassical gravity limit, both perturbative and non-perturbative contributions to the gravitational path integral can be calculated by standard methods.  When a single geometric saddlepoint dominates, this becomes ordinary effective field theory in curved space, and in certain cases with enough supersymmetry, even the sum over geometric saddles can be performed and matched to CFT \cite{Dijkgraaf:2000fq,Dabholkar:2011ec}.

The CFT correlators appearing on the right-hand side of \eqref{dict} are also naturally written as sums, not over field configurations but over conformal blocks. For example, the vacuum correlator $G = \langle O_1 O_2 O_3 O_4\rangle$ can be decomposed into conformal blocks as
\be\label{cftsum}
G = \sum_{\mbox{\footnotesize primaries\ }p} \langle O_1 O_2 \lVert p \rVert O_3 O_4\rangle
\ee
where $\lVert p\rVert $ denotes the projection onto a primary state $p$ and all of its conformal descendants. This sum over blocks must reproduce the bulk path integral, but the map from one to the other is remarkably intricate and understood only in certain limits. In perturbation theory, the mapping from conformal block sums to bulk Witten diagram calculations has been explored extensively in $d$ dimensions; see for example \cite{Heemskerk:2009pn,Hijano:2015zsa}.  A salient feature of this story is that the bulk calculation is always manifestly crossing invariant, since it involves a sum over channels. The CFT calculation, of course, is not manifestly crossing invariant, but crossing-symmetric expansions in the CFT appear to be in one-to-one correspondence with consistent effective field theories in the bulk.

In 3d gravity, this mapping from boundary conformal blocks to bulk calculations can be explored even at the non-perturbative level, in certain cases. In situations where the gravitational backreaction is large, but other interactions are small, the picture that has emerged is that the full nonlinear gravity answer can be reproduced by the Virasoro identity block in CFT \cite{Yin:2007gv,Headrick:2010zt,Hartman:2013mia}. The identity block in two dimensions includes the contributions of all operators built from the stress tensor, so this is an obvious guess --- the all-orders contribution of multiple stress tensors should reproduce nonlinear interactions of the graviton --- but what makes it useful is that technology from Liouville CFT enables one to calculate interesting correlators in great detail and generality, essentially because these Liouville CFT techniques only depend on the conformal algebra. Applications include entanglement entropy \cite{Hartman:2013mia}, thermodynamics \cite{Hartman:2014oaa}, black hole correlators \cite{Fitzpatrick:2014vua}, the information paradox \cite{Fitzpatrick:2016ive}, and collapsing black holes \cite{Anous:2016kss}.

The Virasoro identity block is not unique. It depends on a choice of channel, specifying where to cut the CFT path integral to project onto intermediate states. In all of the applications mentioned above, the working assumption is that the leading gravity answer is equal to the Virasoro identity block in the channel where it is largest:
\be\label{smax}
e^{-S_{\text{bulk}}} \approx \max_\Gamma \left| \mathcal{F}_0^{\Gamma} \right|^2
\ee
where $\mathcal{F}_0^{\Gamma}$ is the Virasoro identity block in the channel $\Gamma$. It is squared to account for left and right movers.  Thus at leading order, on the gravity side we have the bulk action, evaluated on the dominant semiclassical saddle, and in CFT, the identity contribution in the dominant channel. The approximation \eqref{smax} can be completely justified for the thermal partition function  \cite{Hartman:2014oaa} and certain correlators \cite{Kraus:2017kyl} assuming the CFT has a sufficiently sparse spectrum, but in general, it is an assumption, to be tested by comparison to the bulk.

What about other contributions to the path integral in the semi-classical limit? A natural interpretation of \eqref{smax}  is that this is the leading term in the schematic relation
\be\label{ssum}
\sum_{\mbox{\footnotesize saddles}} e^{-S_{\text{bulk}}[g]} = \sum_{\Gamma} \left|\mathcal{F}_0^{\Gamma}\right|^2 \, ,
\ee
and that individual terms on the left are in one-to-one correspondence with individual terms on the right. That is, the gravitational path integral in the semiclassical limit is a sum over channels of the Virasoro identity block, and saddles correspond to channels. This general idea was first introduced with the `black hole Farey tale' \cite{Dijkgraaf:2000fq,MaloneyWitten}, where the thermal partition function is formulated as a sum over modular images, and has since appeared in a variety of contexts. For example, it was applied to correlation functions in \cite{Maloney:2016kee}, and in perturbation theory, it is automatically implemented by Witten diagrams and by the Mellin space formulation of conformal correlators \cite{Fitzpatrick:2011ia}. However, aside from supersymmetric or perturbative examples, it has been difficult to access more than a single leading term in the sum \eqref{ssum}. Our goal is to explore \eqref{ssum} in a simplified setting where the sum over bulk configurations is the sum over worldlines of a single massive particle, moving on a fixed geometry, and the subdominant terms can be probed quantitatively.

The sum \eqref{ssum} agrees with the maximum \eqref{smax} when a single, real saddle dominates. It also concords with the point of view that the gravitational contributions are summarized by an effective Liouville field \cite{Verlinde:1989ua,Coussaert:1995zp, Krasnov:2000zq,Jackson:2014nla}. On the other hand, it is surprising from a CFT point of view, where we normally sum operators in a given, fixed channel, rather than summing over channels. This is justified if, to leading order, the identity operators in various channels do not overlap when dualized into a single channel.

In previous work on large-$c$ correlators, the difference between \eqref{smax} and \eqref{ssum} was purely a matter of interpretation. The leading semiclassical answer was always dominated by a single configuration, and there was no way to test the non-perturbatively suppressed other channels. This will always be the case in Euclidean signature: the bulk action is real, and to leading order, summing a real exponential is equivalent to taking its maximum. But having a  sum, rather than a maximum, is essential in order to interpret the CFT calculation as a bulk path integral, and subleading contributions are physically relevant for questions like late-time behavior \cite{Chen:2016cms,Fitzpatrick:2016mjq}, bulk reconstruction \cite{Balasubramanian:2014sra,Balasubramanian:2016xho}, and extremal CFTs \cite{MaloneyWitten}.

In this paper, we study the 2-point function of a light probe operator during a non-equilibrium thermalization process, building on \cite{Anous:2016kss}. The state is dual to a collapsing black hole in AdS$_3$. We find that for general insertions of the probes, the dominant bulk configuration is a complex worldline of the probe particle, which crosses the collapsing shell at a complex value of the boundary coordinate. This is interesting because it makes it possible to distinguish between the maximum \eqref{smax} and the sum \eqref{ssum}. Interestingly the CFT reproduces the bulk only if we sum over channels, confirming \eqref{ssum}. Put differently, the CFT correlator is not dominated by the Virasoro identity block in any one channel; many channels have identity blocks with the same magnitude but different phases, and these must be summed. The sum over channels can be performed by a saddlepoint approximation --- now on the CFT side --- which leads us to introduce a `complexified OPE channel' dual to a corresponding complex saddle in the bulk.\footnote{At timelike separation, the bulk worldline is always complex, in the sense that the radial coordinate is complex at the turning point. The important difference in the Vaidya case is that the crossing point is also complex in the direction parallel to the boundary, so that the CFT channel also becomes complex.}

\subsection{Setup and summary}
\begin{figure}[t!]
  \begin{centering}
   \includegraphics[width=0.5\textwidth]{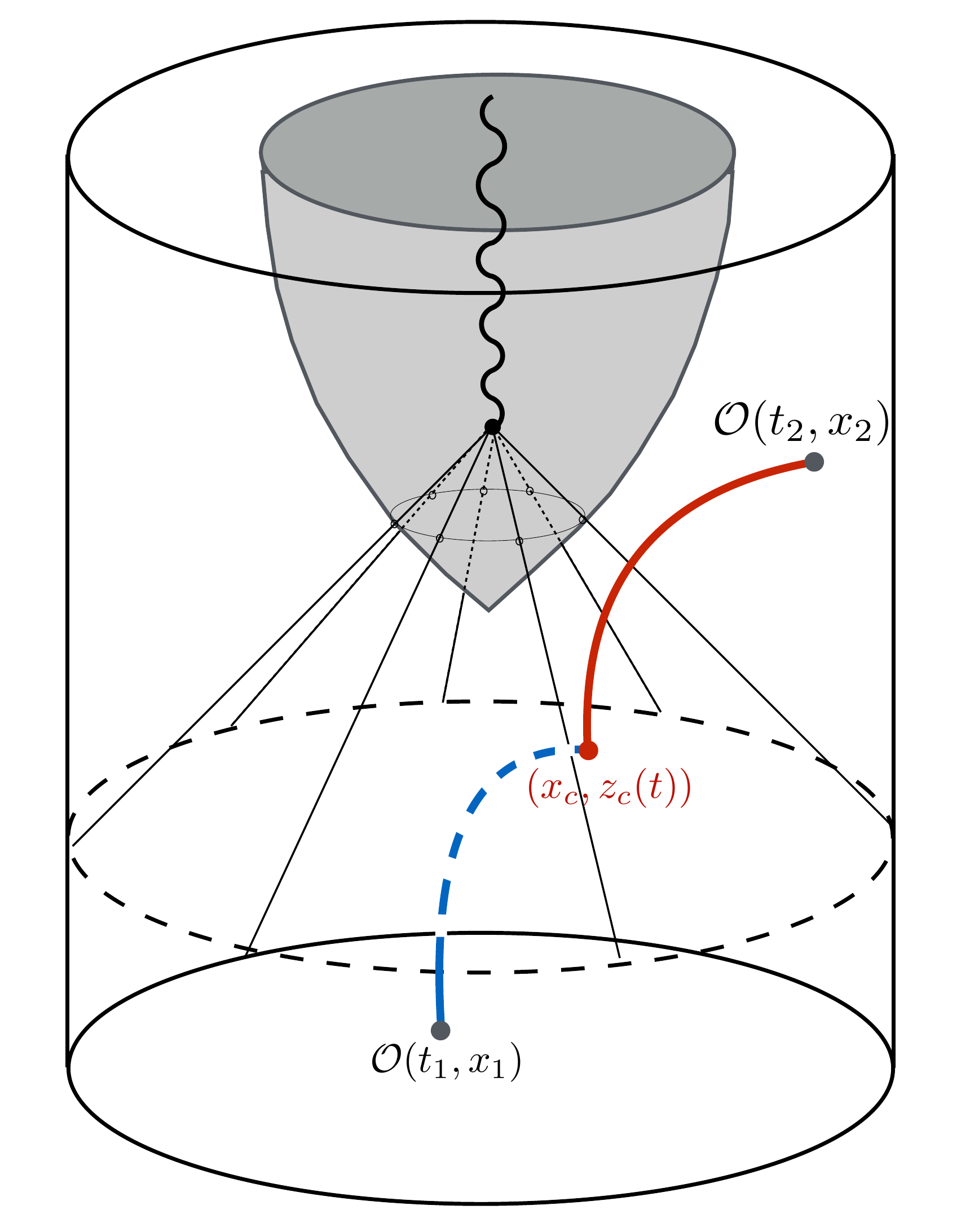}
  \caption{
  Schematic representation of the setup and main result. In a collapsing black hole, the boundary conformal block expansion becomes a sum over  channels labelled by a boundary point $x_c$. This corresponds semiclassically to a bulk geodesic crossing the shall of infalling matter at a point $(x_c,z_c(t))$ with $z_c$ the radial coordinate. Both in the CFT and in the bulk this crossing point takes on complex values, signaling that a complex saddle point dominates the bulk path integral, and no single channel dominates in CFT.  }\label{fig:schema}
\end{centering}
\end{figure}
In more detail, we consider a state $|{\cal V}\rangle$ created by a product of a large number of a local operator insertions at $t=0$. Each operator insertion can be interpreted as creating a highly boosted dust particle in the bulk, so this state is dual to the Vaidya geometry, which describes a collapsing shell of null, pressureless perfect fluid \cite{Anous:2016kss}. In bulk language, the 2-point function of a probe operator is computed by the worldline path integral of a point particle in this background:
\be
 \int Dx(\tau)\,  e^{i m\!\int \! \d\tau} \sim \langle {\cal V}| \cO(x_1) \cO(x_2)| {\cal V}\rangle
\ee
where $m$ is the mass of the particle dual to the operator $\cO$, and the bulk paths $x(\tau)$ are anchored to $x_{1,2}$ at the boundary.  This path integral is a simple case of \eqref{dict}, where the bulk geometry is fixed, but nontrivial, and away from the collapsing shell, the only matter in the bulk is a single point particle. It can be further simplified by splitting the wordline $x(\tau)$ into two (or more) segments, before and after it crosses the collapsing shell. This reduces the bulk path integral to an ordinary integral over the crossing point $x_c$,
\be\label{xcint}
\int \d x_c\, \exp\left( i m L(x_1, x_c) + i m L(x_c, x_2) \right) \sim \langle
\mathcal{V}| \cO(x_1) \cO(x_2)| \mathcal{V}\rangle
\ee
where $L$ is the length of a (potentially complex) geodesic in a Vaidya background. (This is written for a single crossing point $x_c$, but the generalization to multiple crossings is straightforward). Note that despite the appearance of geodesic lengths, this still contains off-shell contributions, due to the integral over $x_c$.

As we will show, each choice of crossing point $x_c$ corresponds to a channel of the boundary OPE, so we may label these channels $\Gamma(x_c)$. We will show that the off-shell worldline labelled by $x_c$ gives a contribution to the bulk path integral equal to the identity block in the corresponding CFT channel:
\be
e^{i m L(x_c)} \approx \left|\mathcal{F}_0^{\Gamma(x_c)}\right|^2 \ .
\ee
This off-shell equality, illustrated in figure \ref{fig:schema}, directly maps the bulk path integral in the form \eqref{xcint} to a sum over channels in the CFT, including subdominant contributions. Performing the sum over CFT channels by a saddlepoint approximation must of course reproduce the bulk, since it is precisely the same sum. When the saddlepoint is real, the sum is dominated by a single channel -- this was the case in \cite{Anous:2016kss}, where we considered the Vaidya geometry with probe operators separated in space or time, but not both. When the saddlepoint is complex, a large family of channels contributes, and they must be summed to reproduce the gravity result. This sum over channels, reinterpreted in a fixed channel, is a sum over heavy operators, so this effectively continuous family of heavy operators is playing an essential role in reproducing thermalization in the bulk.  By summing over channels, we have assumed that the heavy operators corresponding to the identity propagating in each channel are independent from each other. This assumption implicitly restricts the light spectrum and OPE coefficients along the lines of \cite{Hartman:2014oaa,Kraus:2017kyl}.

\section{CFT correlators in the Vaidya state}

\subsection{The state}
In CFT, the Vaidya state on the real line is defined by inserting heavy `dust' operators $\psi$, offset in imaginary time \cite{Anous:2016kss}:
\be
{\cal V} = \prod_{k=-\infty}^\infty \psi(z_k, \bz_k) ,
\quad
z_k = k/n -i \sigma  \ ,
\ee
with $0<\sigma \ll 1$, and the state is $|{\cal V}\rangle = {\cal V}|0\rangle$. (Note that we are quantizing on fixed $\text{Im}\, z$ slices, not radially. Hermitian conjugation acts by reflecting across the real line, so the operators defining $\langle \mathcal{V}|$ are inserted at $z_k^*=k/n+i\sigma$.) We study this state in the limit of large central charge $c \to \infty$ and a large density of insertions, $n \to \infty$. In order to produce a black hole with finite energy density, the limits are taken with energy/(length$\times c$) held fixed, or in terms of the scaling dimension, $h_\psi \sim \sigma c/n$.

In this state, we consider the two-point function of a probe operator ${\cal O}$,
\be\label{gccc}
G(t_1,x_1|t_2,x_2) = \langle {\cal V}| \cO(t_1,x_1) \cO(t_2,x_2)| {\cal V}\rangle \ ,
\ee
where the dimension of ${\cal O}$ satisfies
\be
1 \ll h_\mathcal{O} \ll c \ .
\ee
(We return to the question of operator ordering below.)
All of these limits are designed to accomplish three things. First, the classical geometry is the Vaidya solution, describing a collapsing BTZ black hole. Second, the dual of ${\cal O}$ is a probe particle, massive enough to travel on a geodesic but light enough so that its backreaction can be neglected. And third, the operator dimensions are scaled in a way that enables us to take advantage of a large-$c$ methods in CFT. In particular, the Virasoro conformal blocks simplify dramatically in the large-$c$ limit \cite{bpz,zamo}:
\be\label{expof}
{\cal F}(c, h, \Delta) \approx e^{-\frac{c}{6}f(h/c, \Delta/c)} \ ,
\ee
where $h$ and $\Delta$ are the external and internal dimensions. The semiclassical block $f$ appearing in the exponential can be computed by solving a monodromy problem. As described in \cite{Anous:2016kss}, the monodromy method can be implemented even in the limit of an infinite number of operator insertions, and in the Vaidya state, this renders the calculation tractable as the background becomes translation invariant.

The state $|{\cal V}\rangle$ can also be defined for the CFT on a circle, but inserting operators symmetrically around the circle \cite{Anous:2016kss}. Here we will focus on the CFT on $\mathbb{R}$ for simplicity, but the calculation is easily generalized to the CFT on $S^1$. Formulas in the latter case are presented without derivation in appendix~\ref{ap:circ}.

\subsection{Monodromy prescription for the vacuum block}
To illustrate the discussion in the introduction, we will compute the correlator $G$ with  $t_1<0<t_2$, using large-$c$ CFT methods. The spatial Fourier transform of this correlator was computed via bulk methods in \cite{Keranen:2014lna,Keranen:2015mqc,David:2015xqa}. Since we study the correlator in the real space representation our results and their implications are new in the bulk, while our CFT calculations are entirely new.

Following our notation in \cite{Anous:2016kss}, the large-$c$ two-point function can be obtained by studying the monodromy properties of the differential equation
\begin{equation}\label{eq:mon}
  \chi''(z)+T_{\rm cl}(z)\chi(z)=0
\end{equation}
where
\begin{equation}
  T_{\rm cl}\equiv T_{\textsc h}+\varepsilon\, T_{\textsc l}
\end{equation}
is the expectation value of the stress tensor, which we have split into contributions coming from the heavy insertions defining the Vaidya quench, i.e.\ the state $|{\cal V}\rangle$, and from the light insertions coming from the probe operators $\mathcal{O}$  whose correlation function we are interested in. The small quantity $\varepsilon\equiv 6 h_{\mathcal{O}}/c$, where $h_{\mathcal{O}}$ is the holomorphic weight of $\mathcal O$. To correctly define the Vaidya state --- dual to a collapsing planar-black hole in AdS$_3$ --- we take
\begin{equation}
  T_{\textsc h}(z)= -\frac{\pi^2}{\beta^2}\Theta\big(\text{Im}(z)-\sigma\big)\Theta\big(\text{Im}(z)+\sigma\big)
\end{equation}
with $0<\sigma\ll 1$ as above. Since $T_{\textsc h}(z)$ is holomorphic only away from the $\text{Im}\, z= \pm \sigma$ lines, we will need to supplement the normal monodromy procedure with additional ingredients. We will give a quick review of the procedure in what follows but refer the reader to \cite{Anous:2016kss} for a more in-depth analysis.

The light stress tensor
\begin{equation}
  T_{\textsc l}(z)=\frac{1}{(z-z_1)^2}+\frac{1}{(z-z_2)^2}-\frac{b_1}{z-z_1}-\frac{b_2}{z-z_2}
\end{equation}
has parameters $b_i$ which are fixed by imposing certain monodromy conditions on (\ref{eq:mon}). The basic statement of the monodromy method is that once the $b_i$ are determined, the semiclassical block $f$ appearing in \eqref{expof} can  be calculated from
\be\label{pzf}
\partial_{z_i}f=\frac{6h_{\mathcal{O}}}{c}b_i \, .
\ee
This will eventually allow us to obtain the correlator.

\begin{figure}[t!]
  \begin{centering}
   \includegraphics[width=0.8\textwidth]{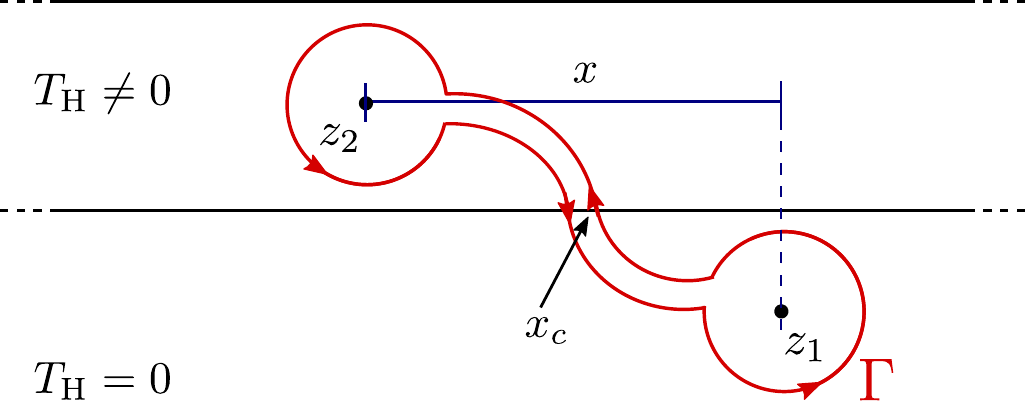}
  \caption{ Monodromy path $\Gamma$ labelled by the crossing point $x_c$.  }\label{fig:crossing}
\end{centering}
\end{figure}

Let $V=(v_1,v_2)$ be a basis of solutions to (\ref{eq:mon}) at $O(\varepsilon^0)$, then at $O(\varepsilon^1)$ the solutions can be written as
\begin{equation}\label{eq:correctedSol}
  \chi(z)=\left(\mathbb{I}+\varepsilon\int^z F\right)\cdot V(z)
\end{equation}
where $F$ is a $2\times 2$ matrix with components
\begin{equation}
  F_i^{~j}=\frac{v_i\epsilon^{jk}v_k}{v_1\,v_2'-v_2\,v_1'}T_{\textsc l}
\end{equation}
where a prime denotes derivation with respect to $z$ and the path used in the integral in \eqref{eq:correctedSol} will be specified later.
We will take $z_1$ outside the strip where $T_{\textsc h}=0$ and $z_2$ inside the strip where $T_{\textsc h}\neq 0$, as shown in figure \ref{fig:crossing}.  Our choice of operator location is the Euclidean analog of placing them respectively before and after the Vaidya quench. In the holographic dual, this means we place the insertions respectively before and after the dust supporting Vaidya is released from the boundary. If we view our CFT procedure as a Euclidean path integral preparing the Vaidya dual, then the insertion $z_1$ placed in the region where $T_{\textsc h}=0$, which upon analytic continuation to Lorentzian signature, captures the information that the CFT is in its vacuum state before the quench.

A basis of solutions to (\ref{eq:mon}) inside and outside the strip are:
\begin{equation}
  V_{\rm inside}=\left(e^{-\frac{\pi z}{\beta}},e^{\frac{\pi z}{\beta}}\right)\, ,\quad\quad\quad\quad V_{\rm outside}=(1,z)\, .
\end{equation}
As is explained in \cite{Anous:2016kss}, we can deal with the discontinuities of $T_{\textsc h}$ by using the jump matrix $J(x_{c})$ defined as follows:
\begin{equation}
  V_{\rm inside}(x_c)= J(x_c) V_{\rm outside}(x_c)\, ,\quad\quad J(x_c)=\frac{1}{\beta}\begin{pmatrix}~~~~(\pi\,x_c+\beta)e^{-\frac{\pi x_c}{\beta}} & -\pi e^{-\frac{\pi x_c}{\beta}}\\-(\pi\,x_c-\beta)e^{\frac{\pi x_c}{\beta}} & \pi e^{\frac{\pi x_c}{\beta}}\end{pmatrix}\, .
\end{equation}
The crossing point $x_c$, see figure \ref{fig:crossing}, labels the path we take in the complex plane to bring the two operators together in the OPE expansion, hence it is a continuous label for the OPE channel. We fix the $b_i$ by demanding that the monodromy matrix $M$ that takes the solutions to (\ref{eq:mon}) around a loop encircling $z_1$ and $z_2$ and crossing at $x_c$ be trivial, that is $M=1$. At first order in $\varepsilon$, this yields the equation
\begin{equation}
\text{Res}_{z_2}F_{\rm inside}+J(x_c)\text{Res}_{z_1}F_{\rm outside}J^{-1}(x_c)  =0~.
\end{equation}
This is a matrix equation for the accessory parameters $b_i$. The solution gives the semiclassical block $f$ in the channel $\Gamma(x_c)$ via \eqref{pzf}:
\begin{equation}\label{eq:semiclassblock}
  f=\frac{12 h_{\mathcal O}}{c}\log\left[\frac{\beta}{\pi}\sinh\left(\frac{\pi(z_2-x_c)}{\beta}\right)-(z_1-x_c)\cosh\left(\frac{\pi(z_2-x_c)}{\beta}\right)\right]~.
\end{equation}
 In integrating \eqref{pzf}, the integration constant is chosen such that $f$ exhibits the correct UV singularities, and the result \eqref{eq:semiclassblock} is given up to an additive constant that is irrelevant because we will compute only the exponential dependence of the correlator.

The contribution to the correlator from a particular conformal block is a product of left- and right-movers,
\be\label{eq:euccor}
G_{\Gamma} (z_{1},z_{2})\approx
 \exp\left(-\frac{c}{6}f(z_1,z_2)-\frac{c}{6}\bar{f}(\bar{z}_1,\bar{z}_2)\right)\, .
\ee
The subscript $\Gamma$ indicates that this is the contribution to the correlator from the vacuum block in the channel $\Gamma=\Gamma(x_c)$.
Let us now analytically continue to Lorentzian time. In general, this can be subtle due to the presence of branch cuts, but here we achieve this simply by performing the following replacements in \eqref{eq:euccor}:
\begin{equation}\label{eq:lorcont}
z_1\rightarrow x-t_1~,\quad \bar{z}_1 \rightarrow  x+t_1~, \quad z_2\rightarrow -t_2~,\quad \bar{z}_2\rightarrow t_2~,
\end{equation}
where we have set $x_2 = 0$ without loss of generality. The analytic continuation of (\ref{eq:euccor}) leads to (with $\Delta\equiv2 h_{\mathcal{O}}$)
\begin{multline}\label{eq:corr}
G_{\Gamma(x_c)}(t_1,x|t_2)=i^{-2\Delta}\left\lbrace\left[\frac{\beta}{\pi}\sinh\left(\frac{\pi(t_2+x_c)}{\beta}\right)-\left(t_1-(x-x_c)\right)\cosh\left(\frac{\pi(t_2+x_c)}{\beta}\right)\right]\right.\times\\\left.\left[\frac{\beta}{\pi}\sinh\left(\frac{\pi(t_2-x_c)}{\beta}\right)-\left(t_1+(x-x_c)\right)\cosh\left(\frac{\pi(t_2-x_c)}{\beta}\right)\right]\right\rbrace^{-\Delta} \ .
\end{multline}
This is the final answer for the contribution of the vacuum representation, in the channel $\Gamma(x_c)$, to the correlator \eqref{gccc}. It is accurate to leading exponential order in $1/c$.

\subsection{Computing the correlator}

The full 2-point correlator is, in principle, given by the vacuum block \eqref{eq:corr} plus the sum over heavy primaries in the channel $\Gamma(x_c)$. By crossing, this produces the same answer for any real value of $x_c$. This holds even when the ${\cal O}$'s are inserted in Lorentzian signature, provided that in regimes where the sum diverges, it is defined by analytic continuation in $z_{1}, \bz_1, z_2, \bz_2$. Note, however, that the label $x_c$ is a choice of channel, not the coordinate of any operator insertion, so even if the operators are inserted at Lorentzian points, $x_c$ is always real and fixed in the usual formulation of the conformal block expansion.

As discussed in the introduction, to reproduce gravity in Euclidean signature, we would choose $x_c$ as the channel where the identity contribution is maximized. In this dominant channel, the full gravity answer is reproduced by the identity block, and heavy operators are suppressed.  This procedure, however, fails in Lorentzian signature, because for real $(x,t_1,t_2)$, the right-hand side of \eqref{eq:corr} is an unbounded function of $x_c$ --- it diverges at one or more points along the real-$x_c$ line.  As we will demonstrate in section \ref{sec:bulkcalc}, the gravity answer is finite except at the expected lightcone singularity.

The resolution of this puzzle is that heavy operators in the conformal block expansion must either cancel, or contribute significantly, in these Lorentzian kinematics. We will show that both possibilities are realized.  When the heavy operators cancel, the gravity result is reproduced by a channel $\Gamma(x_c)$ that extremizes, rather than maximizes, the identity contribution. When the heavy operators become important, they serve to effectively shift the value of $x_c$ into the complex plane.

The first step is to replace the conformal block expansion by a sum over channels:\footnote{While the integrand in (\ref{grep}) may appear to be real, by $\left| \mathcal{F}_0\right|^2$ we simply mean a product over left and right movers $\left| \mathcal{F}_0\right|^2=\mathcal{F}_0\bar{\mathcal{F}}_0$. In Lorentzian signature this product is not real due to $\bar{z}_i\neq z_i^*$. See e.g.\ (\ref{eq:lorcont}) and (\ref{eq:corr}). }
\bea\label{grep}
G &=& \sum_{{\rm primary\ }O_p} \left| \mathcal{F}_p^{\Gamma(x_c)}\right|^2 \notag\\
&\approx& \int_{-\infty}^{\infty} \text{d}x_c \left| \mathcal{F}_0^{\Gamma(x_c)}\right|^2
\eea
where $\left| \mathcal{F}_0^{\Gamma(x_c)}\right|^2 \approx G_{\Gamma(x_c)}(t_1,x|t_2)$ is given in \eqref{eq:corr} at leading order in $1/c$.
That is, instead of summing over all operators in a fixed channel, we will sum the identity block over all channels.  This makes precise the schematic equation \eqref{ssum} discussed in the introduction, adapted to the present context.  In making this replacement, we are assuming that (i) other heavy operators in the theory, which do not correspond to the identity in any channel, are suppressed; and (ii), there is no overlap of the identity in different channels, so that we are not overcounting heavy operators. The first assumption is plausible in a theory with a large gap in operator dimensions above the identity, as in holographic theories. The second assumption is certainly true for any two channels: the identity block in one channel, when reinterpreted in another channel, only has very heavy contributions (in holographic language, above the black hole threshold) \cite{Ponsot:2000mt}. It is less clear for an infinite sum of channels but we will assume that it is true, and view the match with gravity as strong evidence in favor of this proposal. This is similar in spirit to \cite{Maloney:2016kee}.

Performing the integral (\ref{grep}) requires an $i\epsilon$-prescription. This will ensure that the integral is finite, by moving any would-be divergences of the integrand (\ref{eq:corr}) off the real $x_c$ axis. The specific choice of $i\epsilon$-prescription also fixes the time ordering of the resulting correlation function (see section 3 of \cite{Hartman:2015lfa} for a review). Essentially, the ordering in Euclidean time becomes the ordering of operators upon evolving to timelike separation. We will consider the ordering
\begin{equation}\label{eq:order}
  G=\langle\mathcal{V}^\dagger\,\mathcal{O}(t_2)\,\mathcal{V}\,\mathcal{O}(t_1,x)\rangle~.
\end{equation}
This is the choice most amenable to the monodromy prescription, since it corresponds to analytic continuation of operators inserted as shown in figure \ref{fig:crossing}. In the Lorentzian expression \eqref{grep}, the ordering \eqref{eq:order} is achieved by sending $t_1 \to t_1 + i \epsilon$.

Although it is an integral over the real line, the resulting saddlepoint can of course land at a complex value of $x_c$. This corresponds, in practice, to allowing complexified channels in the conformal block expansion, and evaluating the identity block at the extremum rather than the maximum. When the extremal channel has complex $x_c$, it means physically that there is no actual OPE channel where the identity operator dominates --- heavy operators contribute at leading order in any particular channel, but in such a way as to simply shift $x_c$ off the real axis.

We will discuss the subtleties associated with this extremization shortly. The extremization condition means we choose an $x_c$ that solves:
\begin{equation}\label{eq:cond}
{\pi\text{coth}\left(\frac{\pi\left(t_2-x_c\right)}{\beta}\right)-\frac{\beta}{t_1+(x-x_c)}}={\pi\text{coth}\left(\frac{\pi\left(t_2+x_c\right)}{\beta}\right)-\frac{\beta}{t_1-(x-x_c)}}~.
\end{equation}
This equation can have zero, one or several real solutions for $x_{c}$. We will denote solutions (real or complex) to \eqref{eq:cond} by $x_c^{\star}$. Before we discuss the various possibilities, let us first
verify that a solution to (\ref{eq:cond}) when $x=0$ is simply $x_c^{\star}=0$, reproducing the result obtained in \cite{Balasubramanian:2012tu,Anous:2016kss}:
\begin{equation}\label{eq:origcor}
  G(t_1,x=0|t_2)\approx i^{-2\Delta}\left(\frac{\beta}{\pi}\sinh\left(\frac{\pi\,t_2}{\beta}\right)-t_1\cosh\left(\frac{\pi\,t_2}{\beta}\right)\right)^{-2\Delta}~\cdot
\end{equation}
When $x\not =0$, and if there are several solutions to \eqref{eq:cond}, we define a procedure, outlined in the next section, for selecting the correct $x_c^\star$ that matches the integral (\ref{grep}) given the $i\epsilon$-prescription described above.
The resulting value for the correlator is then obtained by plugging $x_{c}=x_{c}^{\star}$ in the right-hand side of \eqref{eq:corr} and we denote the result by
\be
G_{\star}(t_{1},x|t_{2}) \equiv G_{\Gamma(x_{c}^{\star})}(t_1,x|t_2) \, .
\ee

\subsection{Saddle point analysis}\label{sec:saddles}
\begin{figure}[t!]
  \begin{centering}
   \includegraphics[width=0.31\textwidth]{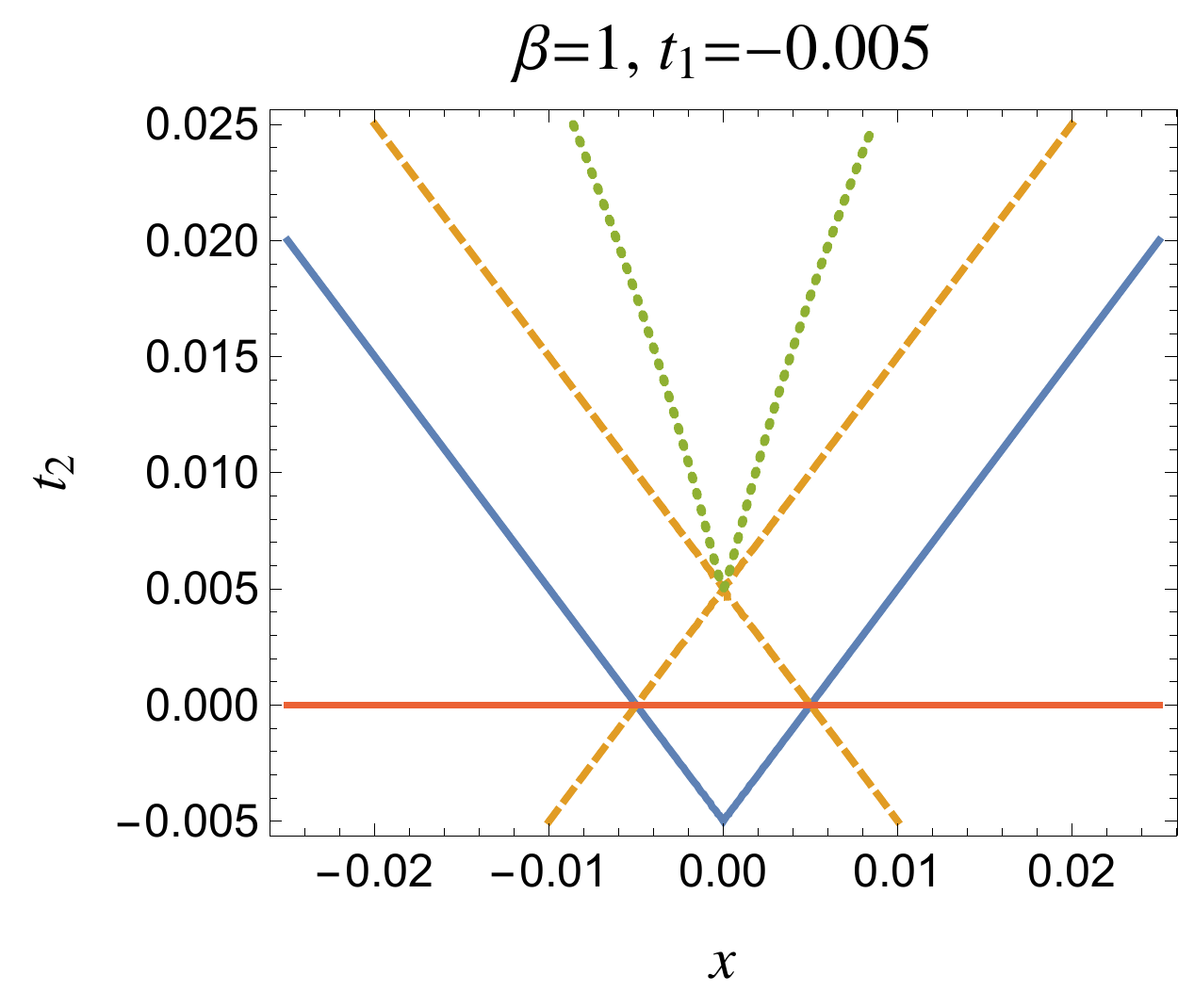}
   \includegraphics[width=0.3\textwidth]{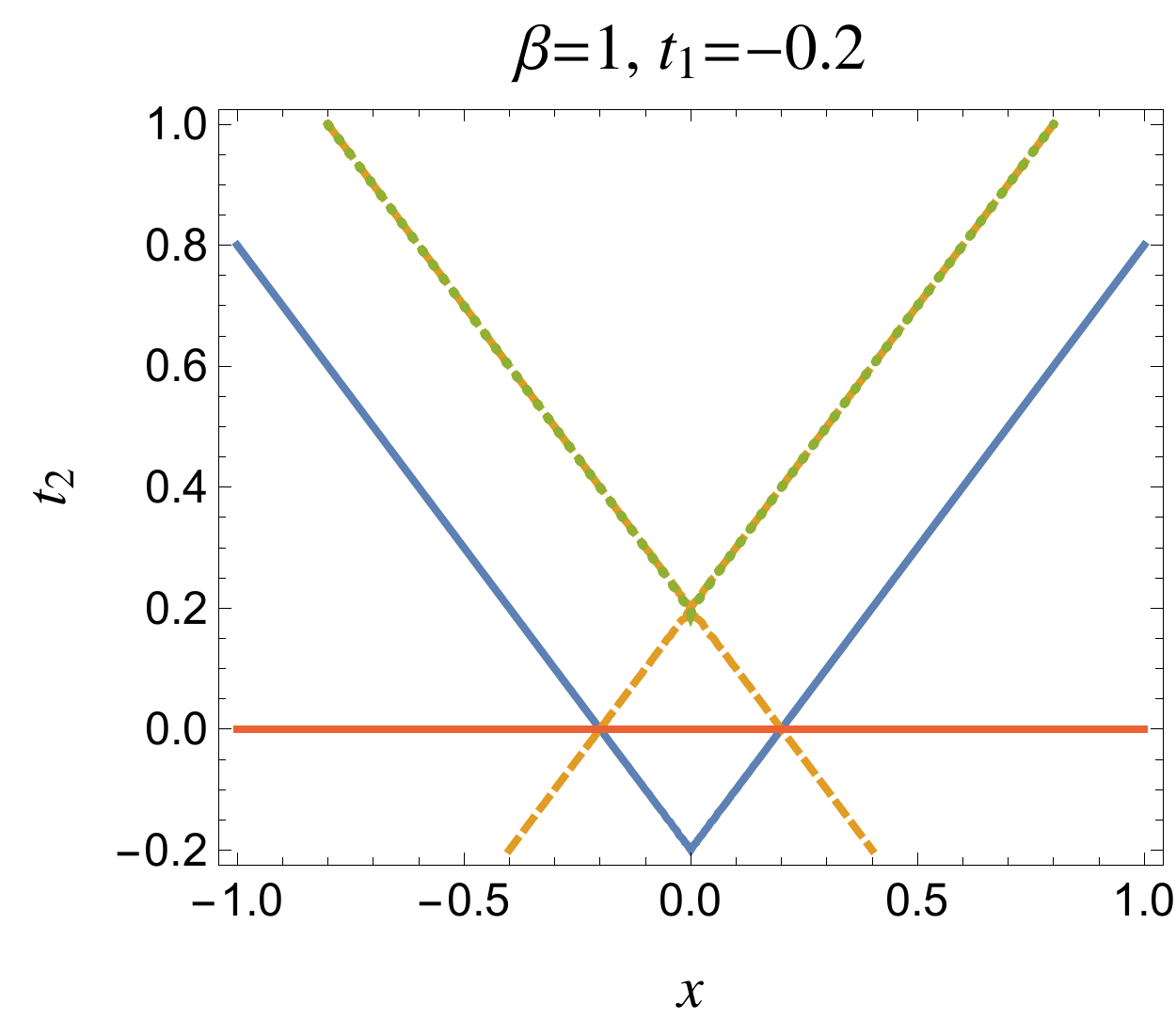}
  \includegraphics[width=0.29\textwidth]{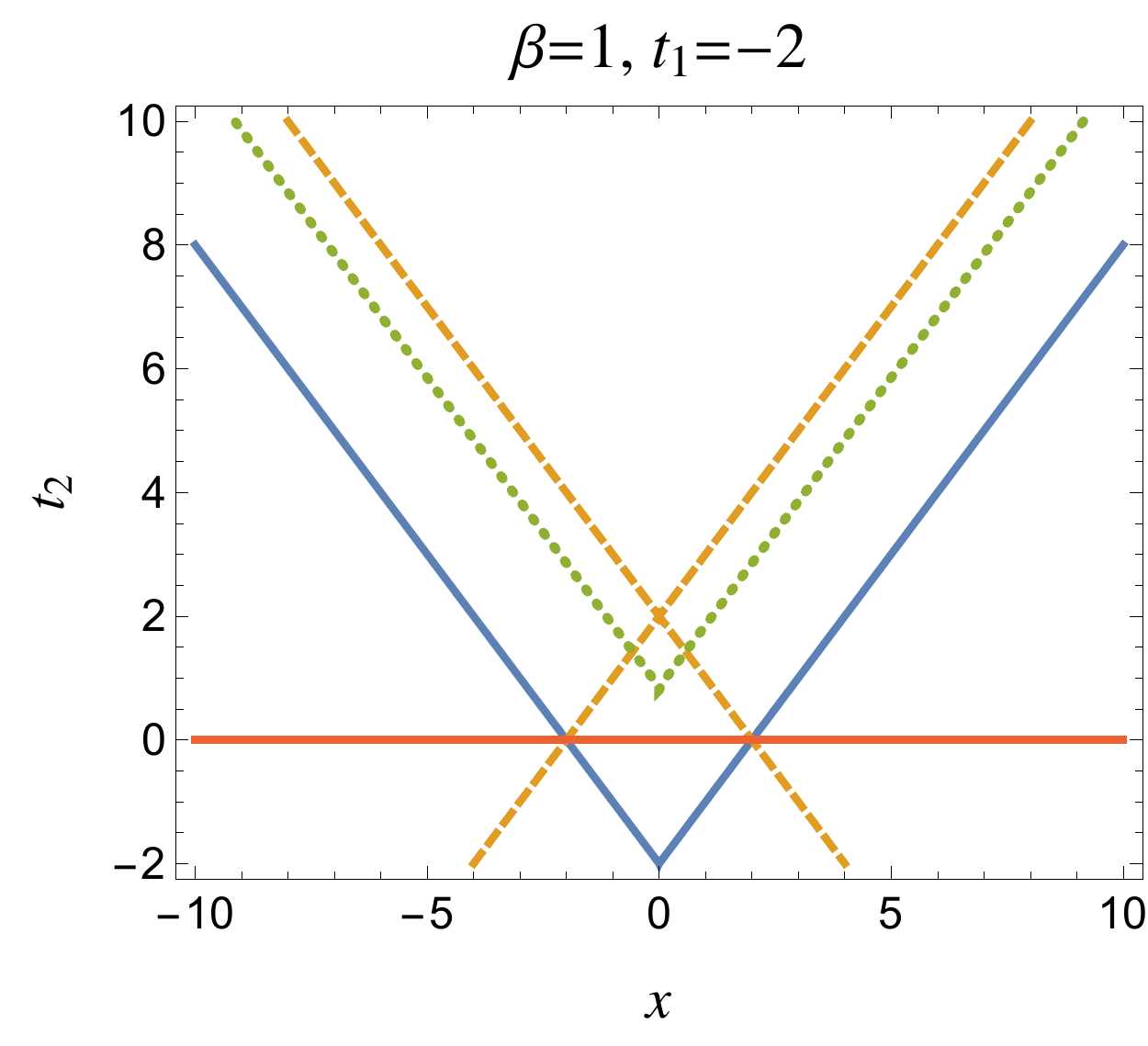}
  \caption{Lines where solutions to the critical equation (\ref{eq:cond}) merge with the real-$x_c$ axis. The solid blue line is the lightcone of $\mathcal{O}(t_1,x)$ defined by $-(t_2-t_1)^2+x^2=0$. The dashed orange line is a mirrored light cone defined by $-(t_2+t_1)^2+x^2=0$. The dotted green line is obtained by evaluating (\ref{eq:t2sol}) at $x_c=y_\star$ with $y_\star$ defined as the solution to (\ref{eq:ystar}). The horizontal red line defines the moment of the quench $t_2=0$.
}\label{fig:regions}
\end{centering}
\end{figure}

The last step is to find the saddlepoint $x_{c}^{\star}$ for a generic Lorentzian configuration of $(x,t_1,t_2)$.
An analytic solution to \eqref{eq:cond} is beyond reach. Instead we can express $G_{\star}(t_1,x|t_2)$ as a parametric function of $x_c$ by solving (\ref{eq:cond}) for $t_2$:
\begin{equation}\label{eq:t2sol}
  t_2=\frac{\beta}{2\pi}\cosh^{-1}\left[\cosh\left(\frac{2\pi x_c}{\beta}\right)-\frac{\pi}{\beta}\frac{t_1^2-(x-x_c)^2}{x-x_c}\sinh\left(\frac{2\pi x_c}{\beta}\right)\right]~.
\end{equation}
We now need to determine the $x_c$ saddle parametrizing $t_2$. In general, there are multiple saddles.  Rather than attempting a detailed analysis of the function in the complex plane, we simply pick the saddle that agrees with numerical integration of \eqref{grep}. The details depend on whether the initial separation at $t_2= 0$ is spacelike or timelike, so we will describe these separately.

Regardless of the location of the insertion $\mathcal{O}(t_1,x)$, equation (\ref{eq:t2sol}) indicates that for $t_2=0$ there always exists a saddle at $x_c^\star=0$. By comparing with the integral (\ref{grep}), and given our choice of $i\epsilon$-prescription, we found that the relevant saddle at $t_2=0$ is always given by $x_c^\star=0$. This corresponds to sitting at the saddle point $x_c^\star=0$ anywhere along the horizontal red lines in figure \ref{fig:regions}.

As we increase $t_2$, the saddle will generically move away from $x_c^\star=0$ along the real-$x_c$ axis until the operator $\mathcal{O}(t_2)$ crosses any one of the three curves depicted in figure \ref{fig:regions}. The solid blue curve is the lightcone of $\mathcal{O}(t_1,x)$ defined by $-(t_2-t_1)^2+x^2=0$. The dashed orange curve is a ``mirrored'' lightcone defined by $-(t_2+t_1)^2+x^2=0$. The dotted green curve is found by evaluating (\ref{eq:t2sol}) at $x_c=y_\star$ with $y_\star$ a solution to:
\begin{equation}\label{eq:ystar}
\tanh\left(\frac{2\pi y_{\star}}{\beta}\right)+\frac{2\pi}{\beta}(x-y_{\star})=0~.
\end{equation}
We will define this time as $t_c\equiv t_2(x_c=y_\star)$.

The initial configuration at $t_2=0$, specifically whether the operators are initially timelike or spacelike separated, determines which of these lines is crossed first, if at all, as $t_2$ is increased. After crossing any one of these three lines, the dominant $x_c^\star$ saddle may move off the real axis into the complex plane or vice-versa. We now proceed to describe the relevant saddles in detail.

\subsubsection*{The case $x=0$}

Let us now revisit the case $x=0$ for which the solution $x_c^\star=0$ is always an exact saddle. In  \cite{Balasubramanian:2012tu,Anous:2016kss} it was assumed that this $x_c^\star=0$ solution is dominant for all configurations $t_1<0<t_2$. We will show that this is not necessarily the case for the ordering \eqref{eq:order}.

\begin{figure}[t!]
  \begin{centering}
   \includegraphics[width=0.3\textwidth]{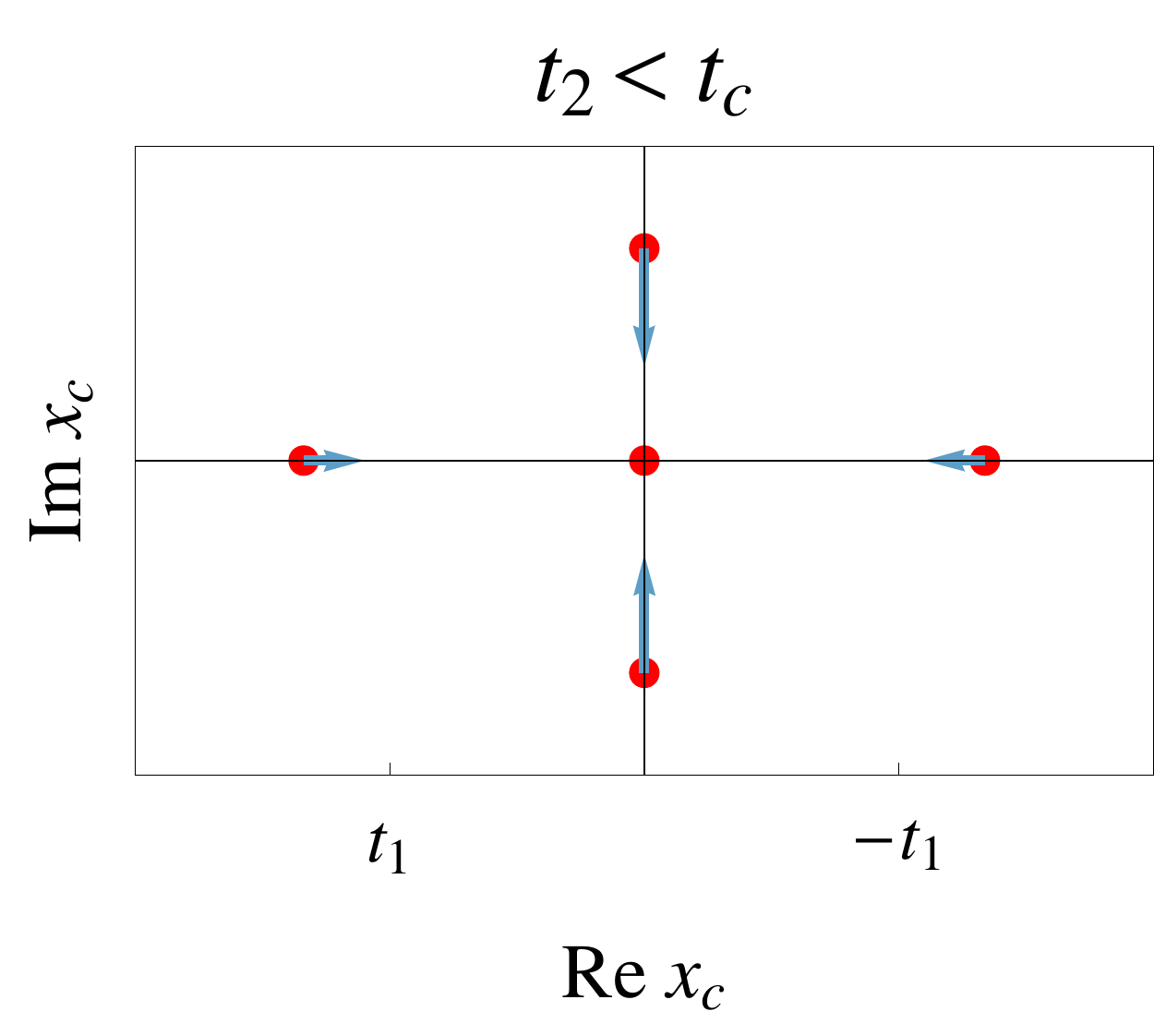}
   \includegraphics[width=0.3\textwidth]{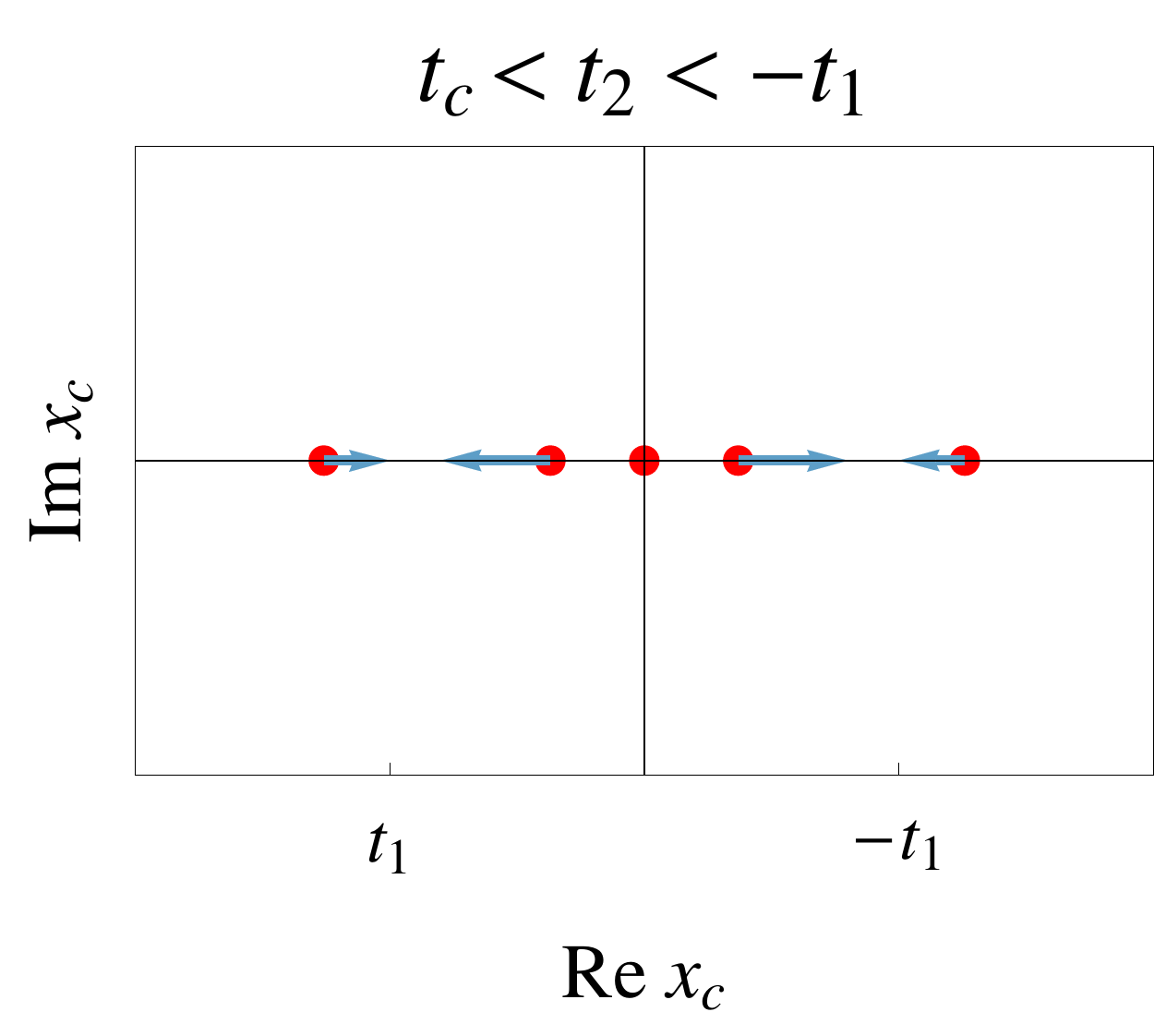}
  \includegraphics[width=0.3\textwidth]{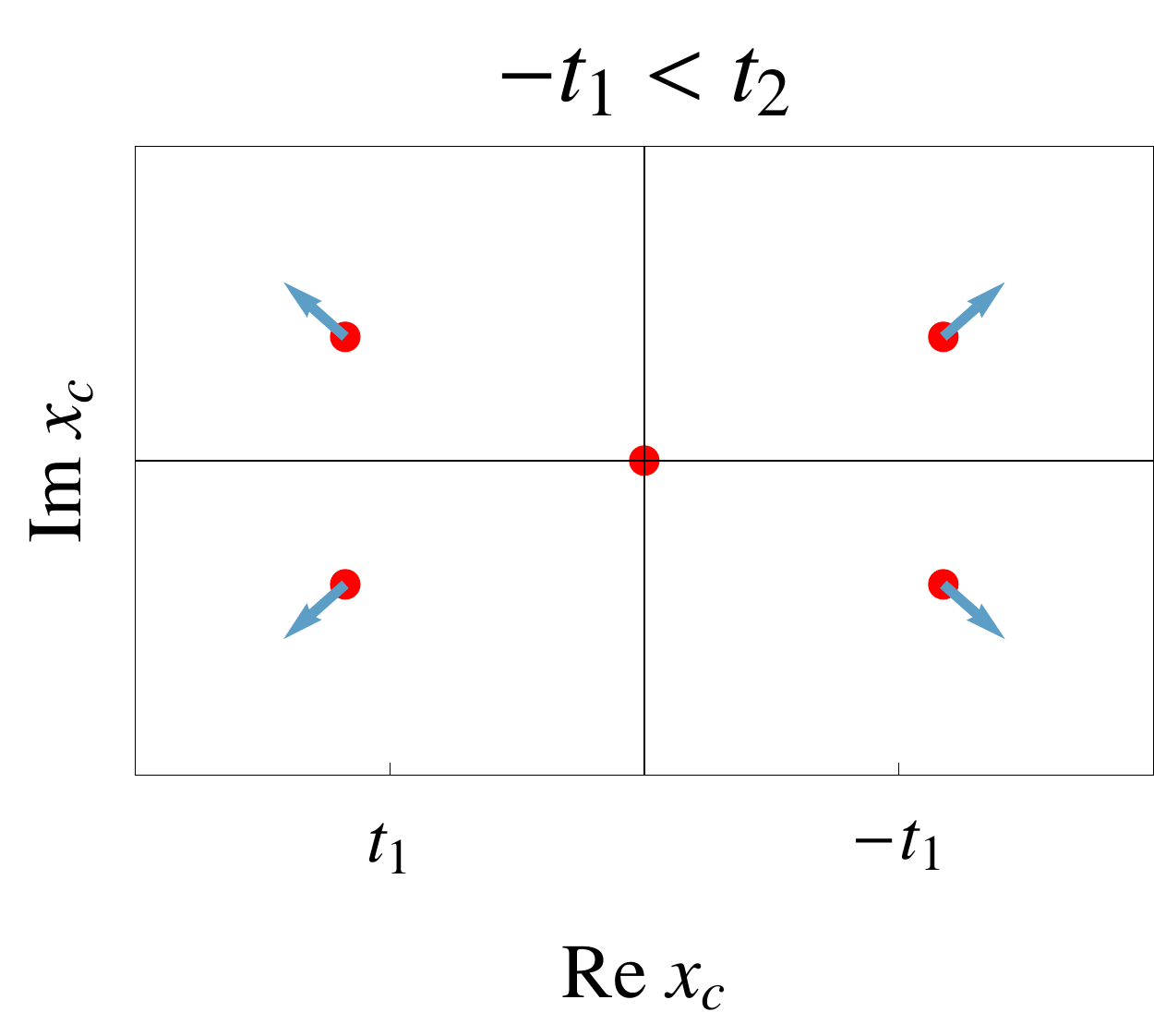}
  \caption{Saddle point solutions to (\ref{eq:cond}) for $x=0$ and fixed $t_1<0$. As $t_2$ increases from zero the number of real solutions goes from three to five to one. Arrows indicate the direction of movement of the saddles for increasing $t_2$.
}\label{fig:saddles}
\end{centering}
\end{figure}

 Let us first describe the saddle points in detail. When $x=0$ the solution to (\ref{eq:ystar}) occurs at $y_\star=0$ and hence the operator $\mathcal{O}(t_2)$ crosses the dotted green curve of figure \ref{fig:regions} at
\begin{equation}
  t_2=t_c=\frac{\beta}{2\pi}\cosh^{-1}\left[1+\frac{1}{2}\left(\frac{2\pi t_1}{\beta}\right)^2\right]<-t_1~.
  \end{equation}
Notice that, as $t_2$ is increased from zero, the dotted green curve is crossed before the ``mirrored" light cone at $t_2=-t_1$. For $t_2<t_c$ there exist three real solutions to (\ref{eq:cond}), including the dominant solution $x_c^\star=0$. As $t_2$ is increased towards $t_2=t_c$, the $x_c^\star=0$ saddle collides with two complex solutions and becomes triply degenerate. As $t_2$ continues to increase for $t_c<t_2<-t_1$ there are five real solutions to (\ref{eq:cond}): the three original real critical points and the two formerly complex solutions that move away from $x_c=0$ along both the positive or negative real-$x_c$ axis. Finally when $\mathcal{O}(t_2)$ crosses the ``mirrored'' lightcone at $t_2=-t_1$, pairs of solutions merge and the three real solutions are: $x_c^\star=0$ and $x_c^\star=\pm t_1$. For $t_2>-t_1$ the additional real solutions move into the complex $x_c$ plane leaving $x_c^\star=0$ as the only real solution. We depict this in pictures in figure \ref{fig:saddles}.

Which of these saddles is picked out by the integral (\ref{grep}) given our $i\epsilon$ procedure? It turns out that it is given by $x_c^\star=0$ for $0<t_2<t_c$, then it moves along the negative real-$x_c$ axis for $t_c<t_2<|t_1|$ until two real solutions merge at $x_c=t_1$ when $\mathcal{O}(t_2)$ crosses the mirrored lightcone of figure \ref{fig:regions}, then the solution moves into the complex plane for $|t_1|<t_2$.

 We will see that this story is basically unchanged for $x>0$ so long as the initial configuration is initially timelike separated, i.e.\ $x^2-t_1^2<0$.

 \subsubsection*{Initially timelike separated: $x^2-t_1^2<0$}

\begin{figure}[t!]
   \begin{centering}
    \includegraphics[width=0.3\textwidth]{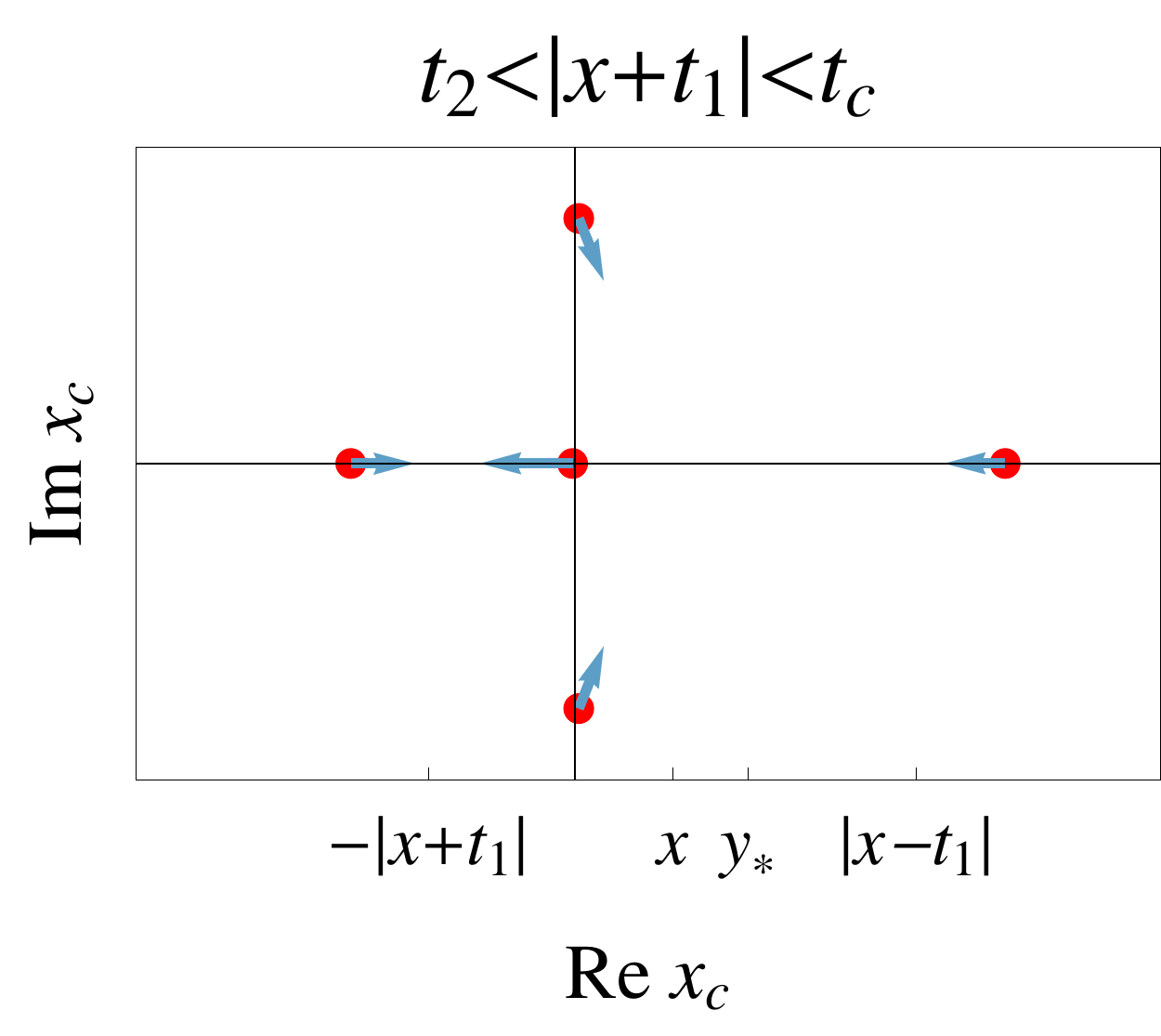}
    \includegraphics[width=0.3\textwidth]{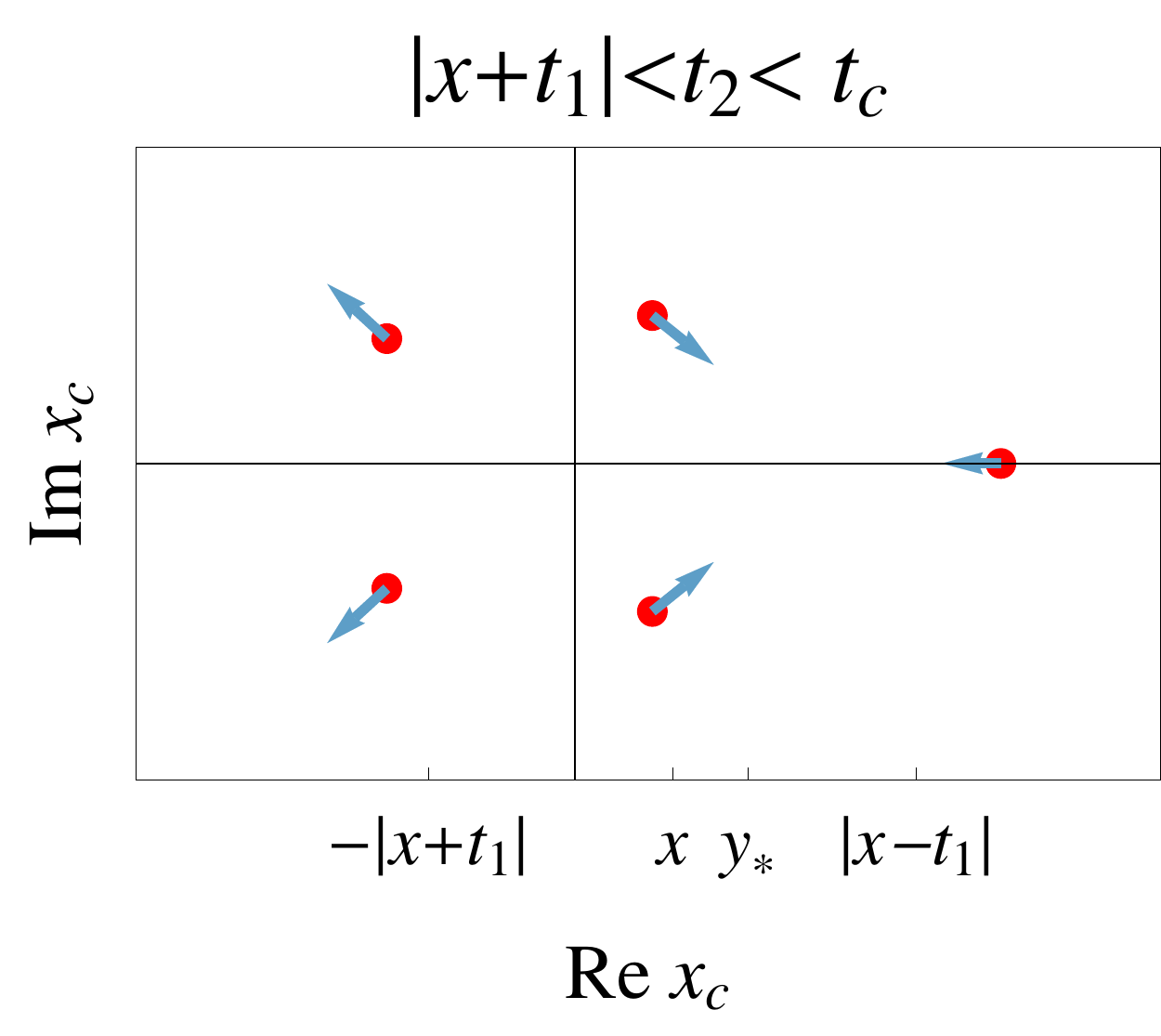}
   \includegraphics[width=0.3\textwidth]{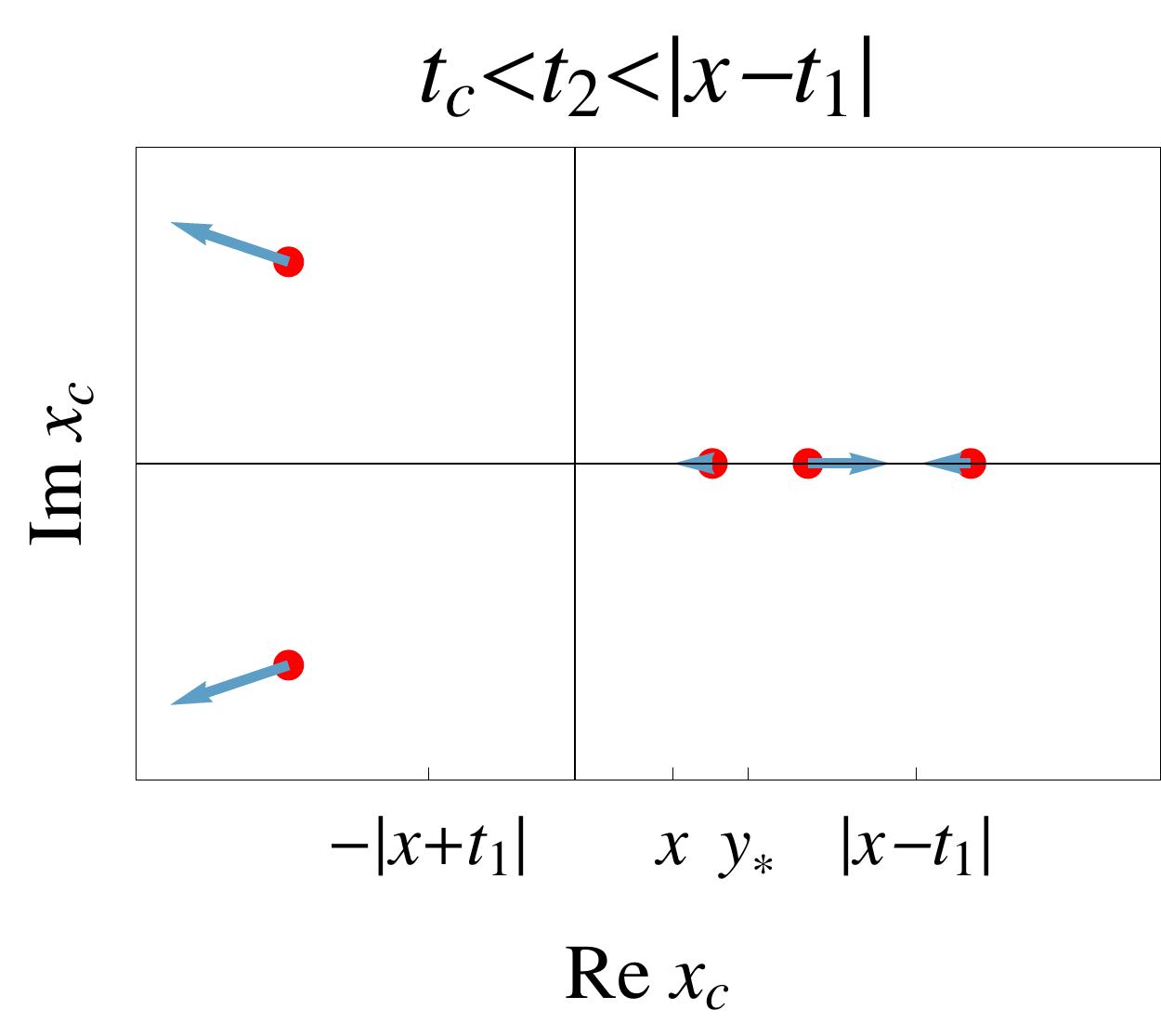}
   \includegraphics[width=0.3\textwidth]{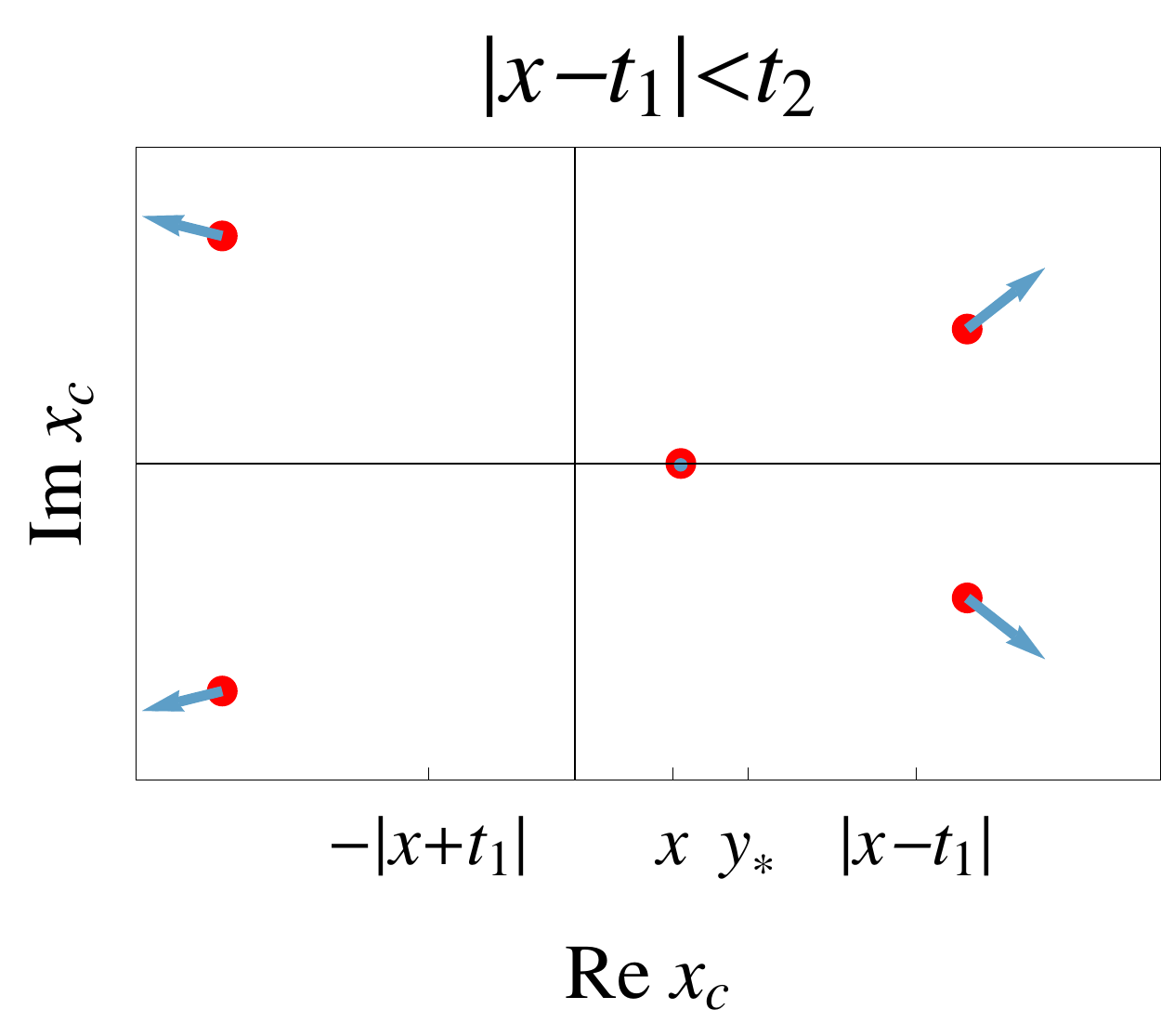}
   \caption{Saddle point solutions to (\ref{eq:cond}) for $0<x<-t_1$ and fixed $t_1<0$. We have picked a configuration where $|x+t_1|<t_c<|x-t_1|$ however the story is similar for any possible ordering.
 }\label{fig:saddlestimelike}
 \end{centering}
 \end{figure}

 For $0<x<-t_1$, the story is analogous to the strict $x=0$ case. However, for $x\neq 0$, the solution $x_c^\star=0$ is only exact when $t_2=0$. As $t_2$ is increased, this solution moves along the negative real-$x_c$ axis until it collides with another real saddle. This happens when $\mathcal{O}(t_2)$ crosses the mirrored lightcone at $t_2=|x+t_1|$, upon which both of these solutions become complex. Unlike the strict $x=0$ case, the original $x_c^\star=0$ solution does not collide with the complex saddles that exist in the range $0<t_2<\text{min}\{t_c,x-t_1\}$. These instead merge with the real-$x_c$ line at $x_c=\text{min}\{y_\star,x-t_1\}$. As $t_2$ continues to increase, one saddle moves left towards $x_c=x$ while the other solution merges with yet another real saddle once $\mathcal{O}(t_2)$ crosses $\text{max}\{t_c,x-t_1\}$ corresponding to either the mirrored lightcone  $t_2=|x-t_1|$ or the $t_2=t_c$ curve, whichever comes first as depicted in firgure \ref{fig:regions}. The motion of these saddles as $t_2$ is increased with $x$ and $t_1$ fixed is presented in figure \ref{fig:saddlestimelike}.
 The strict $x=0$ case described in the previous section can be thought of as a degenerate limit of this discussion wherein $y_\star=x=0$. This implies that some of the special points in figure \ref{fig:saddlestimelike} collapse onto the origin.

 The relevant saddle that matches onto our integral is the one that starts at $x_c^\star=0$ and moves left until it collides with another saddle and then moves into the complex plane. Hence when plotting the correlation function for insertions which are timelike separated at $t_2=0$ we will make parametric plots of $G_\star(t_1,x|t_2)$ with parameter $x_c^\star$ following a contour as in the left plot of figure \ref{fig:t2contours}.

 \subsubsection*{Initially spacelike separated: $x^2-t_1^2>0$}
 \begin{figure}[t!]
   \begin{centering}
    \includegraphics[width=0.3\textwidth]{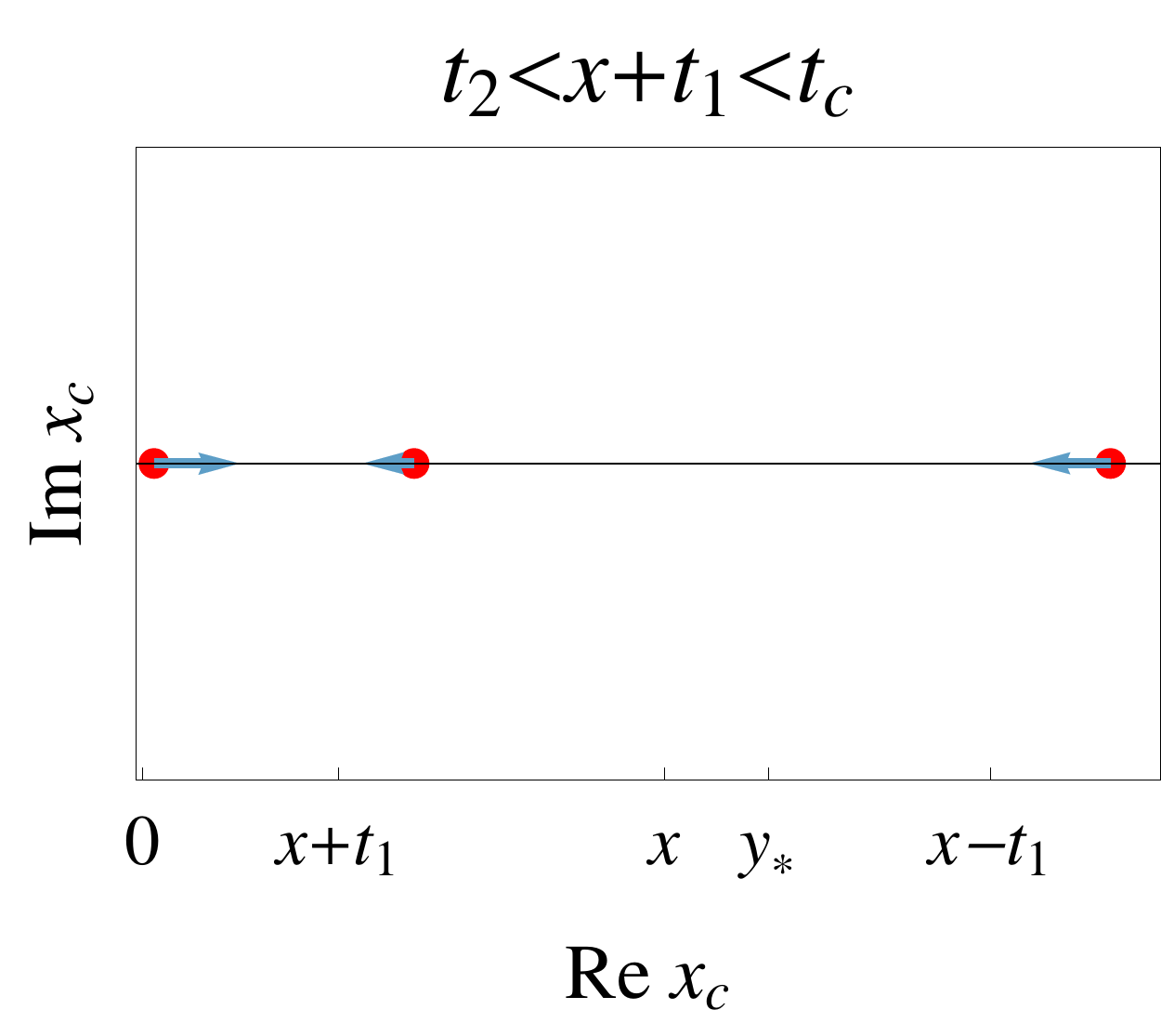}
    \includegraphics[width=0.3\textwidth]{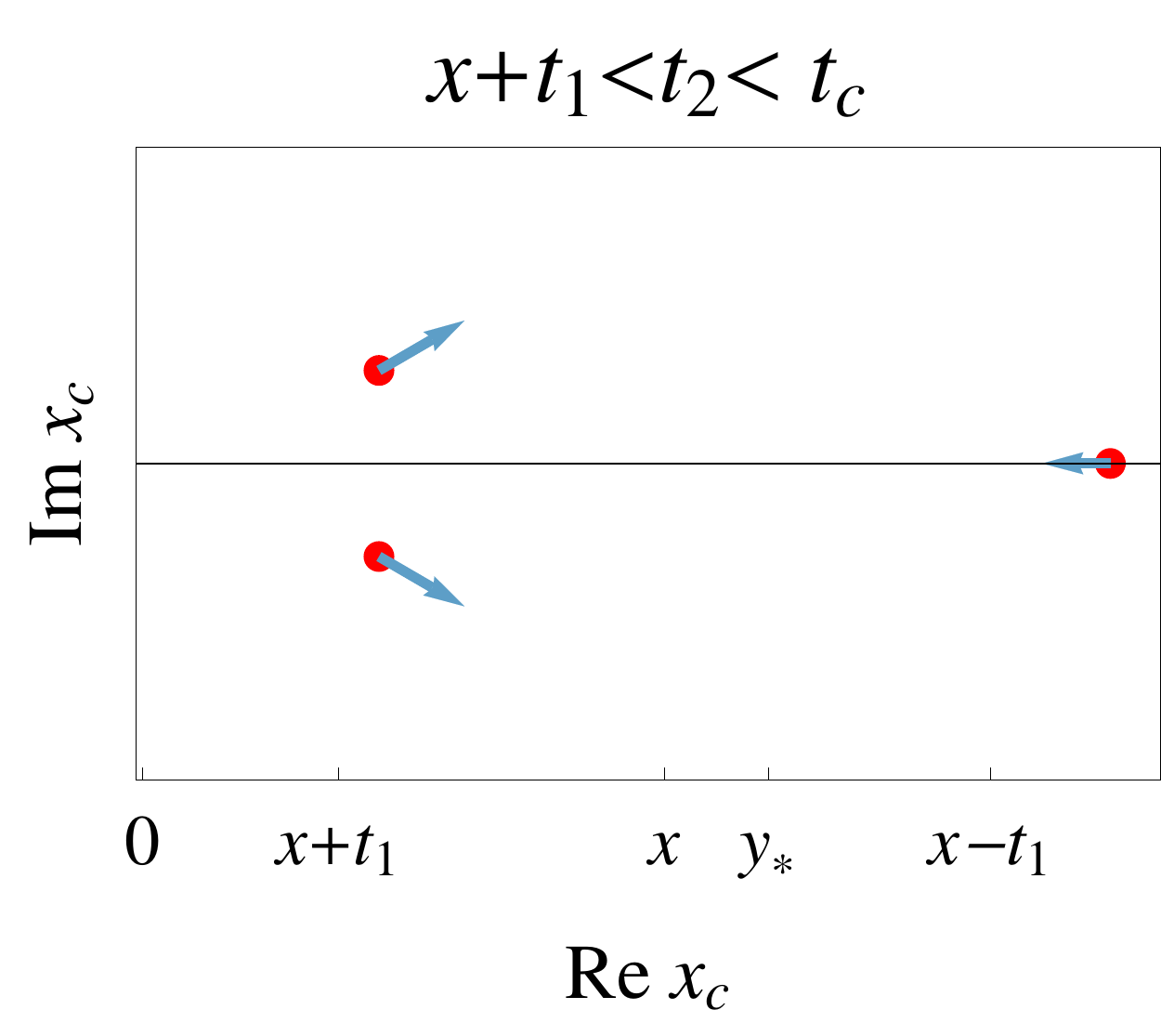}
   \includegraphics[width=0.3\textwidth]{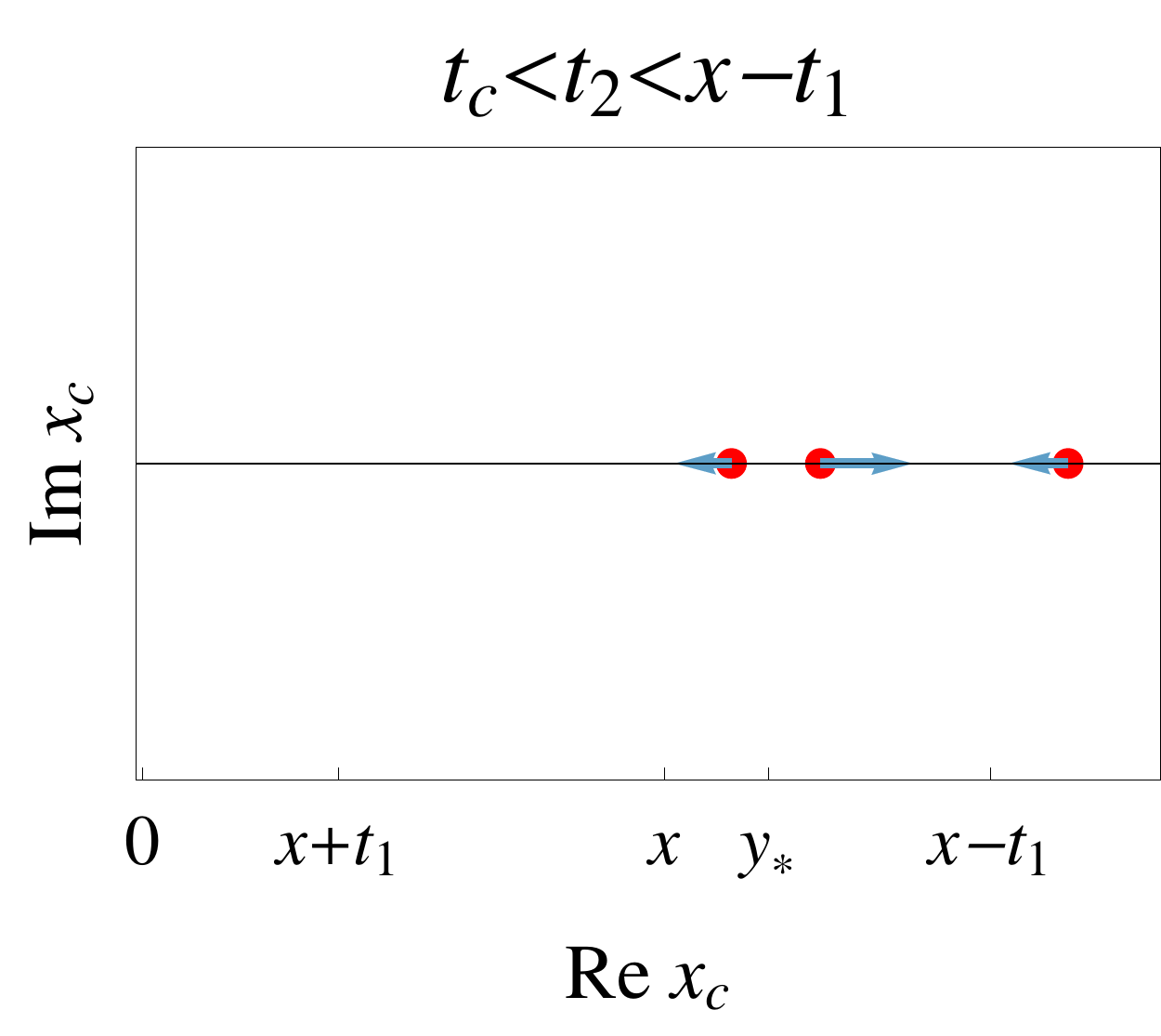}
   \includegraphics[width=0.3\textwidth]{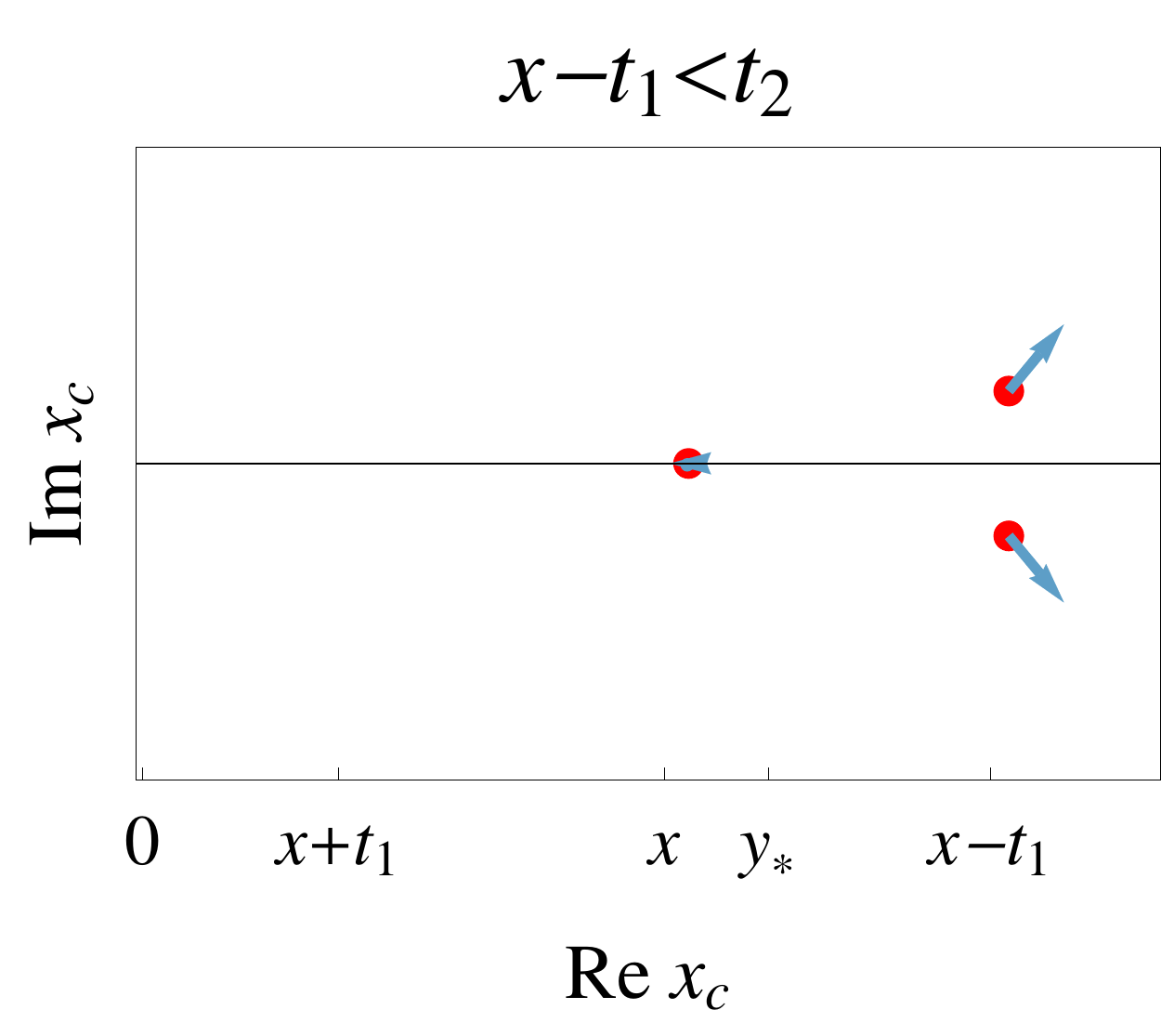}
   \caption{Saddle point solutions to (\ref{eq:cond}) for $-t_1<x$. We have picked a configuration where $t_c<x-t_1$, however the story is exactly the same for $x-t_1<t_c$ with their rolls reversed.
 }\label{fig:saddlesspacelike}
 \end{centering}
 \end{figure}

 For $-t_1<x$ we encounter a slightly different scenario. Here the $x_c^\star=0$ solution is again exact for $t_2=0$, however it now moves to the \emph{right} as we increase $t_2$. Once $\mathcal{O}(t_2)$ crosses the lightcone of $\mathcal{O}(t_1,x)$ this saddle collides with another and they both move into the complex $x_c$ plane. These saddles merge with the real-$x_c$ axis at $x_c=\text{min}\{x-t_1,y_\star\}$ corresponding to $t_2=\text{min}\{x-t_1,t_c\}$. Either case is possible as shown in figure \ref{fig:regions}. Once this happens one of the saddles moves towards $x_c=x$ while the other moves towards $x_c=\text{max}\{x-t_1,y_\star\}$. This signals another collision of saddles where both again become complex for $\text{max}\{x-t_1,t_c\}<t_2$. We depict this in pictures in figure \ref{fig:saddlesspacelike}.

\begin{figure}[t!]
  \begin{centering}
      \includegraphics[width=0.4\textwidth]{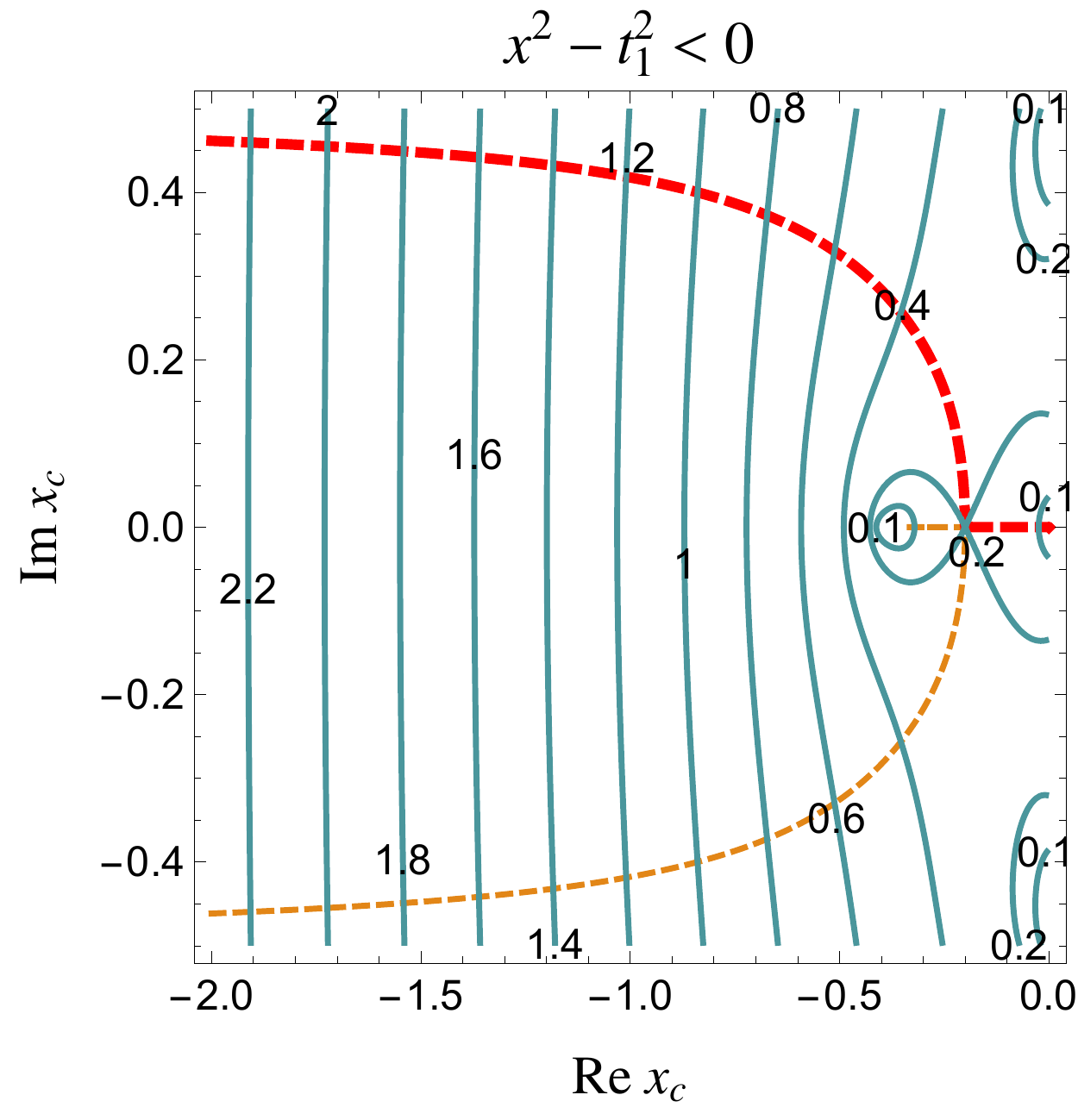}
      \includegraphics[width=0.4\textwidth]{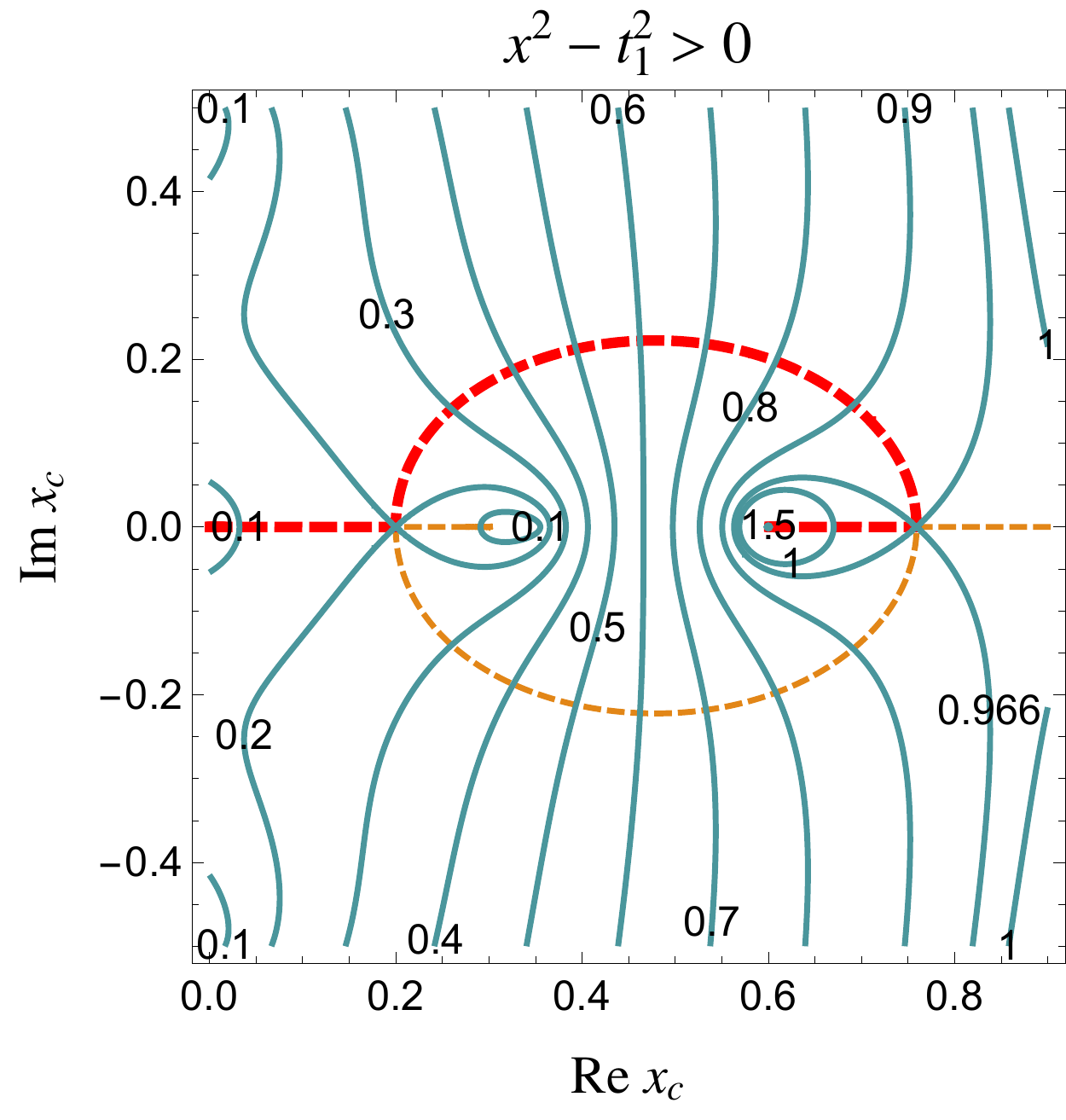}
  \caption{
Motion of saddlepoints in the complex-$x_c$ plane.  The dashed red curve is the saddlepoint $x_c^{\star}(t_2)$, parameterized by real $t_2 \in [0,\infty]$, that solves \eqref{eq:t2sol}.  The solid curves are contours of constant $\text{Re}\,t_2(x_c)$ evaluated on the RHS of \eqref{eq:t2sol} and the dashed curves have $\text{Im}\,t_2(x_c)=0$.
\emph{Left:} Insertions initially timelike separated at $t_2=0$ with $\beta=1$, $x=0.6$, $t_1=-0.8$. The saddlepoint starts at the lower right at $t_2 = 0$, and moves to the left as $t_2$ increases.
\emph{Right:} Insertions initially spacelike separated at $t_2=0$ with $\beta=1$, $x=0.6$,  $t_1=-0.4$. The saddlepoint starts at the origin at $t_2 = 0$, and moves initially to the right as $t_2$ increases.
}\label{fig:t2contours}
\end{centering}
\end{figure}

 Again our integral procedure picks out the $x_c^\star=0$ saddle at early times, which becomes complex after the lightcone singularity. This solution then becomes real and once it merges with its complex conjugate, then moves left or right along the real axis as $t_2 \to \infty$ (which direction is not important for our purposes).   An example of this contour is depicted in the right hand figure of \ref{fig:t2contours}.

 After plotting some example correlation functions in the next section, we will proceed to show that the bulk computation of the same correlator, via a Witten diagram, picks out the \emph{exact same} complex saddles once we specify the correct $i\epsilon$ procedure, this time in the bulk.
 These complex saddles are not mysterious from the bulk perspective, as the Witten diagram involves integrating over a bulk point. However, without (\ref{grep}) we would have no way of interpreting them on the CFT side.

\begin{figure}[h!]
  \begin{centering}
   \includegraphics[width=0.4\textwidth]{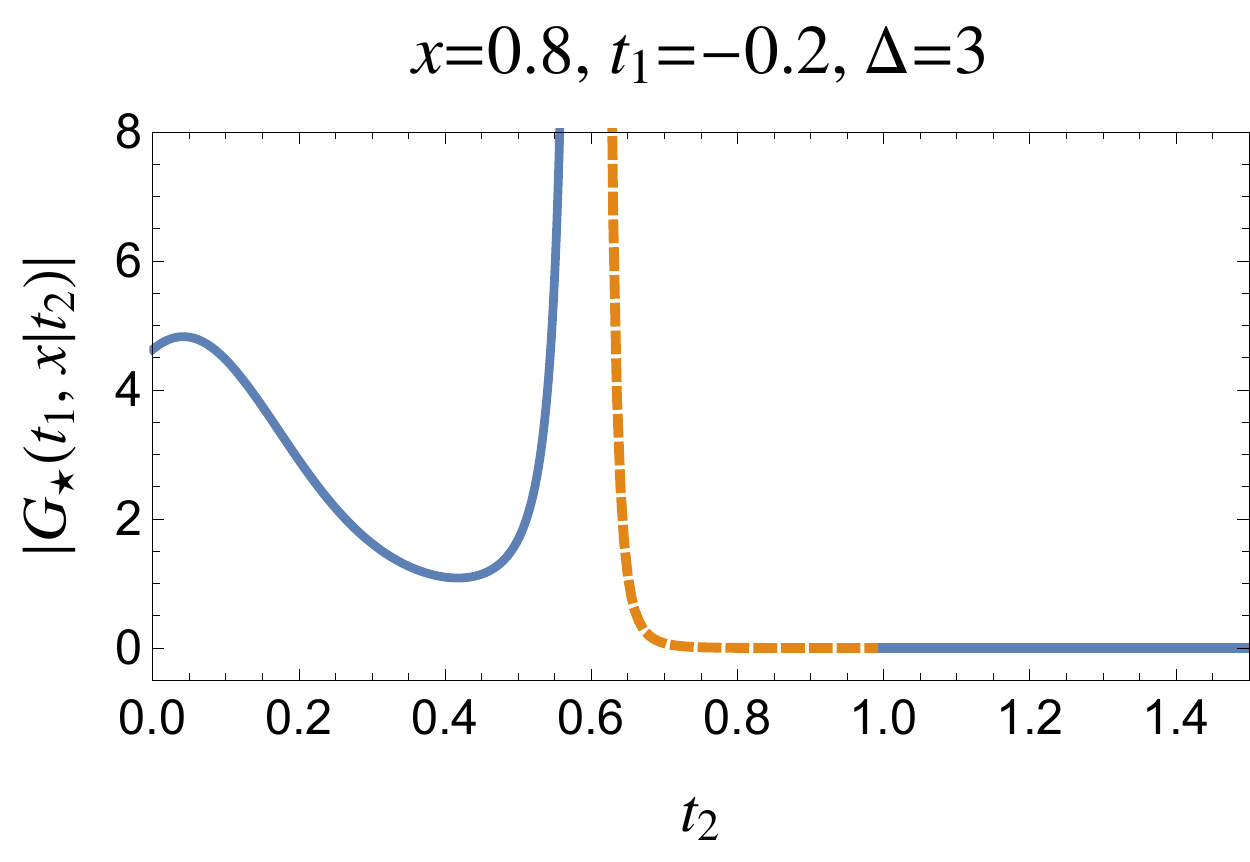}
  \includegraphics[width=0.4\textwidth]{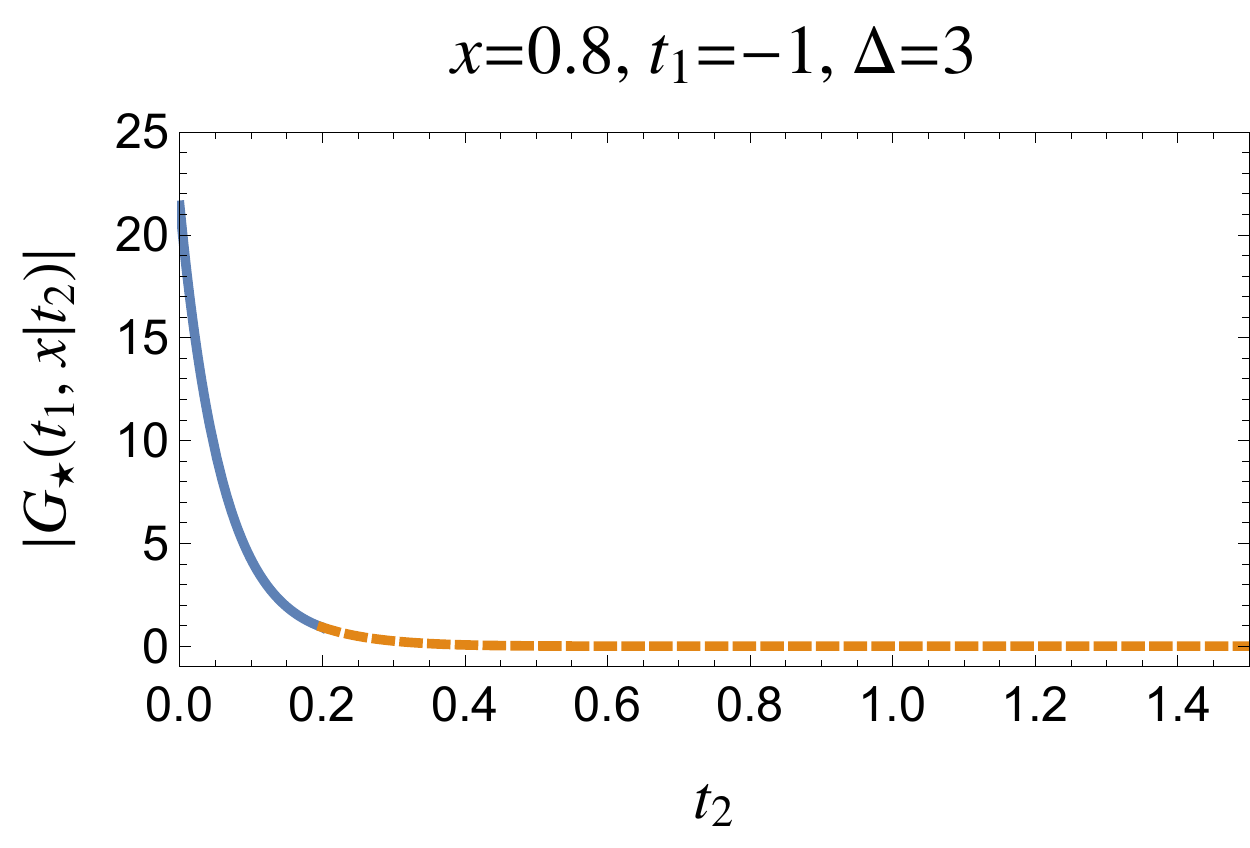}
  \caption{\emph{Left:} Two-point function for insertions initially spacelike separated at $t_2=0$ with $\beta=1$, $x=0.8$,  $t_1=-0.2$ and $\Delta=3$. The divergence is the expected lightcone singularity at $t_2=x+t_1=0.6$~.  \emph{Right:} Two-point function for insertions initially timelike separated at $t_2=0$ with $\beta=1$, $x=0.8$, $t_1=-1$ and $\Delta=3$. The solid blue parts of the curves represent configurations with purely real $x_c^\star$. The orange dashed sections of the curves represent configurations with complex $x_c^\star$.  }\label{fig:corr}
\end{centering}
\end{figure}

\subsection{Plots}

We have analyzed how to represent $G_\star(t_1,x|t_2)$ as a parametric function along a complex $x_c^\star$ contour. These contours are depicted in figure \ref{fig:t2contours} and we plot a few examples of the correlation function in figure \ref{fig:corr}. The correlation function so computed exhibits all expected features, including the lightcone singularity for $t_2=x+t_1>0$, as well as the exponential decay at late times.\footnote{ We note that for the CFT on $\mathbb{R}$, the exponential decay is not in contradiction with unitarity as it would be for the CFT on $S^1$ where it would signify information loss at leading order in the $1/c$ expansion \cite{Anous:2016kss}.} In our plots we distinguish between the portions where $x_c^\star$ is complex versus purely real. Note that, as displayed, the correlation function away from the lightcone singularity is both continuous and smooth, giving further evidence that we have chosen the correct saddles.

\section{Bulk calculation}\label{sec:bulkcalc}

We now proceed to show that our choice of complex $x_c^\star$ has a precise analog in the holographic calculation, leading to a match between correlators on both sides of the duality. We will calculate the same correlation function as in the previous section, but this time by evaluating a Witten diagram in planar-AdS$_3$-Vaidya. The Vaidya metric for an infinitesimally thin shell reads
\begin{equation}
\d s^2=\frac{\ell^2}{z^2}\left(-F(z,v)\d v^2-2\d v \d z+\d x^2\right)~,\quad\quad\quad F(z,v)\equiv 1-\Theta(v)\left(\frac{2\pi z}{\beta}\right)^2~,
\end{equation}
and describes a simple black hole collapse geometry, obtained by gluing vacuum AdS$_3$ to BTZ along the null surface $v=0$. To go back to more standard coordinates in each patch of the spacetime we substitute
\begin{equation}\label{eq:vjoin}
  v=\begin{cases}t-z~, &v<0\\ t- \frac{\beta}{2\pi}\tanh^{-1}\left(\frac{2\pi z}{\beta}\right)~, &v>0~.\end{cases}
\end{equation}
To obtain $G(t_1,x_1|t_2,x_2)$ with $t_1<0<t_2$ and $(t_i,x_i)$ boundary points, we will evaluate the leading Witten diagram. This leading diagram starts at $(t_1,x_1,z=0)$, gets propagated to the null shell using the retarded boundary-to-bulk planar AdS$_3$ propagator, and then from the null shell back to the boundary at $(t_2,x_2,z=0)$ using the retarded bulk-to-boundary planar BTZ propagator. This configuration is depicted in figure \ref{fig:schema}.

The vacuum AdS$_3$ retarded bulk-to-bulk propagator for a scalar of dimension $\Delta$ satisfies
\begin{equation}
  \Big(-\square+\frac{\Delta(\Delta-2)}{\ell^2}\Big)G_{\textsc r}(x,x')=\frac{\delta(x-x')}{\sqrt{-g}}~,
\end{equation}
and vanishes for $t'<t$. By symmetry it must be a function of the chordal distance:
\begin{equation}
  u_{\rm AdS}(x,x')\equiv \frac{-(t-t')^2+(x-x')^2+z^2+z'^2}{2z\,z'}~,
\end{equation}
and must vanish as $z^\Delta$ if $z\rightarrow 0$. This implies
\begin{equation}\label{eq:adsBB}
  G_{\textsc r}^{\rm AdS}=\Theta\left(t'-t\right) u_{\rm AdS}^{-\Delta} \ {}_2{F_1}\left(\frac{\Delta+1}{2},\frac{\Delta}{2},\Delta,u_{\rm AdS}^{-2}\right)
\end{equation}
up to an overall normalization and inclusion of $i\varepsilon$'s. We can extract the boundary-to-Bulk and Bulk-to-boundary propagators using the peeling method (or from Green's theorem):
\begin{align}
  G_{\text{bB}}^{\rm AdS}&\equiv~\lim_{z\rightarrow 0}(2\,z)^{-\Delta} G_{\textsc r}^{\rm AdS}(x,x')=\Theta\left(t'-t\right)\left(\frac{z'}{-(t-t')^2+(x-x')^2+z'^2}\right)^\Delta~,\\
G_{\text{Bb}}^{\rm AdS}&\equiv\lim_{z'\rightarrow 0}(2\,z')^{-\Delta} G_{\textsc r}^{\rm AdS}(x,x')=\Theta\left(t'-t\right)\left(\frac{z}{-(t-t')^2+(x-x')^2+z^2}\right)^\Delta~.
\end{align}
The BTZ analog of these propagators are obtained by starting with (\ref{eq:adsBB}) and replacing $u_{\rm AdS}$ by $u_{\rm BTZ}$ in the right hand side of \eqref{eq:adsBB}, where
\begin{equation}
  u_{\rm BTZ}\equiv\frac{\cosh\left(\frac{2\pi\left(x-x'\right)}{\beta}\right)-\cosh\left(\frac{2\pi\left(t-t'\right)}{\beta}\right)\sqrt{\left(1-\left(\frac{2\pi z}{\beta}\right)^2\right)\left(1-\left(\frac{2\pi z'}{\beta}\right)^2\right)}}{\left(\frac{2\pi z}{\beta}\right)\,\left(\frac{2\pi z'}{\beta}\right)}
\end{equation}
which obeys $\lim_{\beta\rightarrow\infty}u_{\rm BTZ}=u_{\rm AdS}$. Now we want to compute :
\begin{equation}\label{eq:WittenDiag}
  G(t_1,x_1|t_2,x_2)=\int_{-\infty}^{\infty} \d x_c \int_{0}^{\infty} \d z_c \sqrt{-g}\,g^{v\mu}\,\left[ G_{\text{bB}}^{\rm AdS}(t_1,x_1,z_1=0|x_c,z_c)\overleftrightarrow{\partial_\mu} G_{\text{Bb}}^{\rm BTZ}(x_c,z_c|t_2,x_2,z_2=0)\right]_{v=0}
\end{equation}
where $(x_c,z_c)$ is a point on the shell located at $v=0$ and $f\overleftrightarrow{\partial_\mu}g\equiv f\partial_\mu g-(\partial_\mu f)g$. This formula can be proven using Green's identities, the basic idea being that we can think of half of the Vaidya spacetime as the BTZ spacetime with a boundary at $v=0$. An initial condition slightly before the shockwave gets propagated into the BTZ spacetime using a modified version of the above equation, and we treat the AdS propagator as that initial condition. The evaluation at $v=0$ means that we take $v\rightarrow 0^-$ for the AdS propagator and $v\rightarrow 0^+$ for the BTZ propagator, using (\ref{eq:vjoin}).

The integral in (\ref{eq:WittenDiag}) is hard to evaluate in general, but for sufficiently large $\Delta$ it admits a saddle point approximation. To leading order in $\Delta$
\begin{equation}\label{eq:adsSaddle}
  G(t_1,x_1|t_2,x_2)\approx e^{-\Delta S_\star}~,
\end{equation}
where
\begin{align}
  S&\equiv-\lim_{\Delta\rightarrow\infty}\frac{1}{\Delta}\log \bigg( \sqrt{-g}\,g^{v\mu}\,\left[ G_{\text{bB}}^{\rm AdS}(t_1,x_1,z_1=0|x_c,z_c)\overleftrightarrow{\partial_\mu} G_{\text{Bb}}^{\rm BTZ}(x_c,z_c|t_2,x_2,z_2=0)\right]_{v=0} \bigg) \nonumber\\
  &=-\log\left[\frac{4\pi^2 z_c^2\,e^{\frac{2\pi x_c}{\beta}}/\beta}{\left(-t_1^2+(x_c-x_1)^2+2t_1 z_c\right)\left(\beta\left[e^{\frac{4\pi x_c}{\beta}}+e^{\frac{4\pi x_2}{\beta}}\right]-2e^{\frac{2\pi(x_c+x_2)}{\beta}}\left[\beta\cosh\left(\frac{2\pi t_2}{\beta}\right)-2\pi z_c\sinh\left(\frac{2\pi t_2}{\beta}\right)\right]\right)}\right]
\end{align}
and $S_\star$ is evaluated on the solution of
\begin{equation}
  \partial_{x_c}S=\partial_{z_c}S=0\, .
\end{equation}
To check that we have done things correctly, we evaluate (\ref{eq:adsSaddle}) in the simplest case where $x_1=x_2=0$. There is a saddle point at:\footnote{We note, as in the previous section, that this may not necessarily be the dominant saddle for all $(t_1,t_2)$, but still use this as a check of our procedure.}
\begin{equation}
  x_c^{\star}=0~,\quad\quad\quad z_c^{\star}=\frac{t_1}{1+\frac{\pi t_1}{\beta}\,\text{coth}\left(\frac{\pi t_2}{\beta}\right)}
\end{equation}
and we recover
\begin{equation}
  G(t_1,t_2)\approx i^{-2\Delta}\left(\frac{\beta}{\pi}\sinh\left(\frac{\pi\,t_2}{\beta}\right)-t_1\cosh\left(\frac{\pi\,t_2}{\beta}\right)\right)^{-2\Delta}~,
\end{equation}
as expected. This agrees with (\ref{eq:origcor}) obtained using the CFT monodromy method when $x=0$. Moreover, notice that the saddle point value of $x_c$ corresponds precisely to the saddle point value of the crossing point in the CFT calculation. This is no accident. We will now show that this holds true at nonzero spatial separation.

 By translation invariance, the general result will only depend on $x_2-x_1$, hence from now on we will set $x_2=0$ and $x_1=x$. It is straightforward to solve $\partial_{z_c}S=0$ for $z_c$, yielding
\begin{equation}
  z_c^{\star}=\left(\frac{t_1}{t_1^2-(x_c-x)^2}+\frac{\frac{\pi}{\beta}\sinh\left(\frac{2\pi\,t_2}{\beta}\right)}{\cosh\left(\frac{2\pi\,t_2}{\beta}\right)-\cosh\left(\frac{2\pi\,x_c}{\beta}\right)}\right)^{-1}~.
\end{equation}
What remains is to solve
\begin{multline}\label{eq:adscond}
  0=\left(\frac{\cosh\left(\frac{2\pi t_2}{\beta}\right)-\left[\cosh\left(\frac{2\pi x_c}{\beta}\right)-\frac{\pi}{\beta}\frac{t_1^2-(x-x_c)^2}{x-x_c}\sinh\left(\frac{2\pi x_c}{\beta}\right)\right]}{\cosh\left(\frac{2\pi t_2}{\beta}\right)-\cosh\left(\frac{2\pi x_c}{\beta}\right)}\right)\times\\
  \left(\frac{\cosh\left(\frac{2\pi t_2}{\beta}\right)-\left[\cosh\left(\frac{2\pi x_c}{\beta}\right)-\frac{\pi}{\beta}\frac{t_1^2-(x-x_c)^2}{t_1}\sinh\left(\frac{2\pi t_2}{\beta}\right)\right]}{\cosh\left(\frac{2\pi t_2}{\beta}\right)-\left[\cosh\left(\frac{2\pi x_c}{\beta}\right)+\frac{\pi}{\beta}\frac{t_1^2-(x-x_c)^2}{t_1}\sinh\left(\frac{2\pi t_2}{\beta}\right)\right]}\right)
\end{multline}
for $x_c$ and evaluate the correlation function
\begin{equation}\label{eq:adscor}
  G(t_1,x|t_2)=i^{-2\Delta}\left(\frac{2\left(t_1^2-(x_c-x)^2\right)\left[\cosh\left(\frac{2\pi\,t_2}{\beta}\right)-\cosh\left(\frac{2\pi\,x_c}{\beta}\right)\right]}{\left[\left(t_1^2-(x_c-x)^2\right)\sinh\left(\frac{2\pi\,t_2}{\beta}\right)-\frac{t_1\beta}{\pi}\left\lbrace\cosh\left(\frac{2\pi\,t_2}{\beta}\right)-\cosh\left(\frac{2\pi\,x_c}{\beta}\right)\right\rbrace\right]^2}\right)^{\Delta}
\end{equation}
on this extremal value of $x_c$.

Notice that the saddle point equation in CFT (\ref{eq:t2sol}) automatically satisfies (\ref{eq:adscond}). Hence there exists a branch of saddles for which $x_c^\star$ in AdS is in one-to-one correspondence with $x_c^\star$ in CFT. Using the on-shell condition (\ref{eq:cond}) we can massage (\ref{eq:adscor}) such that its expression is exactly that of (\ref{eq:corr}). This establishes that (\ref{eq:adscor}) and (\ref{eq:corr}) are equal once evaluated on the on-shell solution $x_c^\star$ solving (\ref{eq:cond}) and (\ref{eq:adscond}).

The bulk integral (\ref{eq:WittenDiag}) requires an $i\epsilon$-prescription to make it finite and well-defined. This prescription picks out one of the saddles of (\ref{eq:adscond}), and we have shown that one branch of these saddles is in one-to-one correspondence with saddles on the CFT side as described by (\ref{eq:cond}), including the complex saddles described in section \ref{sec:saddles}. Hence the saddle point analysis of the bulk Witten diagram calculation matches precisely with the corresponding analysis of the sum over identity channels of the CFT, confirming the sum prescription (\ref{grep}).

We conclude by emphasizing once more that the complexification of $x_c^\star$ is completely natural from the point of view of the Witten diagram---it  implies that no real configuration dominates the integral (\ref{eq:WittenDiag}) and that the steepest descent curve moves into the complex plane. This leads us to take the same intepretation in CFT, this time viewed as a sum over conformal blocks as in (\ref{grep}).

\section*{Acknowledgments}
We are grateful to Alice Bernamonti, Ben Craps, Federico Galli, Christoph Keller, Alex Maloney, and G\'{a}bor S\'{a}rosi for discussions.
TA is supported by the Natural Sciences and Engineering Research Council of Canada, and by grant 376206 from the Simons Foundation. TH is supported by DOE grant DE-SC0014123. The research of JS and AR is supported by the Fonds National Suisse de
la Recherche Scientifique (FNS) under grant number 200021 162796 and
by the NCCR 51NF40-141869 ``The Mathematics of Physics" (SwissMAP).

\appendix

\section{CFT on \texorpdfstring{$S^1$}{S1}}\label{ap:circ}
The discussion in the main text applies in the more general case of correlators probing a Vaidya quench in a large-$c$ CFT on $S^1$. In this appendix we provide the final formulas without derivation, but the interested reader should find it straightforward to obtain these results using a combination of the methods found in the main text and in \cite{Anous:2016kss}.

The correlation function on the circle of radius $R=1$ with $t_1<0<t_2$ in the Vaidya quench is:
\begin{align}\label{eq:corrcirc}
G(t_1,\theta|t_2)=i^{-2\Delta}\Bigg\lbrace& \left[\frac{\beta}{\pi}\cos\left(\frac{t_1-(\theta-\phi_c)}{2}\right)\sinh\left(\frac{\pi(t_2+\phi_c)}{\beta}\right)-2\sin\left(\frac{t_1-(\theta-\phi_c)}{2}\right)\cosh\left(\frac{\pi(t_2+\phi_c)}{\beta}\right)\right]\times\nonumber\\ &\left[\frac{\beta}{\pi}\cos\left(\frac{t_1+(\theta-\phi_c)}{2}\right)\sinh\left(\frac{\pi(t_2-\phi_c)}{\beta}\right)-2\sin\left(\frac{t_1+(\theta-\phi_c)}{2}\right)\cosh\left(\frac{\pi(t_2-\phi_c)}{\beta}\right)\right]\Bigg\rbrace^{-\Delta}~,
\end{align}
with $\phi_c$ determined by the equation:
\begin{equation}\label{eq:condcirc}
\pi\text{coth}\left(\frac{\pi\left(t_2-\phi_c\right)}{\beta}\right)-\frac{\beta}{2}\cot\left(\frac{t_1+(\theta-\phi_c)}{2}\right)=\pi\text{coth}\left(\frac{\pi\left(t_2+\phi_c\right)}{\beta}\right)-\frac{\beta}{2}\cot\left(\frac{t_1-(\theta-\phi_c)}{2}\right)~.
\end{equation}
We can solve (\ref{eq:condcirc}) for $t_2$:
\begin{equation}\label{eq:t2solcirc}
  t_2=\frac{\beta}{2\pi}\cosh^{-1}\left[\cosh\left(\frac{2\pi \phi_c}{\beta}\right)-\frac{4\pi\sin\left(\frac{t_1+(\theta-\phi_c)}{2}\right)\sin\left(\frac{t_1-(\theta-\phi_c)}{2}\right)}{\beta\sin(\theta-\phi_c)}\sinh\left(\frac{2\pi \phi_c}{\beta}\right)\right]~,
\end{equation}
which allows us to plot $G_\star(t_1,\theta|t_2)$ as a parametric function of $\phi_c$ along a complex contour where $t_2$ is real and monotonically increasing.

\end{spacing}

\bibliographystyle{utphys}
\bibliography{threeDblackholerefs}{}

\end{document}